\documentclass[12pt,thmsb,titlepage]{article}
\usepackage{amsmath}
\usepackage{amssymb}
\usepackage{graphicx}
\usepackage{epsf}
\usepackage[T1]{fontenc}
\usepackage[latin9]{inputenc}
\usepackage[letterpaper]{geometry}
\usepackage{setspace}
\usepackage{epstopdf}

\usepackage{rotating}

\usepackage{epstopdf}

\providecommand{\tabularnewline}{\\}

\usepackage{epsf}
\usepackage[hidelinks]{hyperref}
\usepackage{amsmath}
\usepackage{amssymb}
\usepackage{graphicx}

\usepackage{caption}
\usepackage{subcaption}

\newcommand{\captionfonts}{\footnotesize}
\makeatletter  % Allow the use of @ in command names
\long\def\@makecaption#1#2{%
  \vskip\abovecaptionskip
  \sbox\@tempboxa{{\captionfonts #1: #2}}%
  \ifdim \wd\@tempboxa >\hsize
    {\captionfonts #1: #2\par}
  \else
    \hbox to\hsize{\hfil\box\@tempboxa\hfil}%
  \fi
  \vskip\belowcaptionskip}
\makeatother   % Cancel the effect of \makeatletter

\newtheorem{theorem}{Theorem}

\newtheorem{lemma}{Lemma}
\newtheorem{corollary}{Corollary}
\newtheorem{ass}{Assumption}

\begin{document}

\centerline{\Large{\bf Inference with Many Weak Instruments }}

\medskip

\centerline{Anna Mikusheva\footnote{
Department of Economics, M.I.T. Address: 77 Massachusetts Avenue, E52-526, Cambridge, MA, 02139. Email: amikushe@mit.edu. National Science Foundation support under grant number 1757199 is gratefully acknowledged. We are grateful to Josh Angrist, Kirill Evdokimov, Whitney Newey and Mikkel S{\o}lvsten for advice,  to Brigham Frandsen for sharing code for simulations, and to Ben Deaner and Sylvia Klosin for research assistance.} and Liyang Sun\footnote{UC Berkeley and CEMFI. Email: lsun20@berkeley.edu. Support from the Jerry A. Hausman Graduate Dissertation Fellowship and Institute of Education Sciences, U.S. Department of Education, through Grant R305D200010, is gratefully acknowledged.}}
\medskip

\centerline{\bf Abstract}
\smallskip

\noindent {\small{We develop a concept of weak identification in linear IV models in which the number of instruments
can grow at the same rate or slower than the sample size. We propose a jackknifed version of
the classical weak identification-robust Anderson-Rubin (AR) test statistic. Large-sample
inference based on the jackknifed AR is valid  under heteroscedasticity and weak identification. The feasible version of
this statistic uses a novel variance estimator. The test has uniformly correct size and good power properties.
We also develop a pre-test for weak identification that is related to the size property of a Wald test based on the Jackknife Instrumental Variable  Estimator.  This new pre-test is valid under
heteroscedasticity and with many instruments.
}
\smallskip

\noindent\texttt{Key words:}  instrumental variables, weak identification, dimensionality
asymptotics.
\smallskip

\noindent\texttt{JEL classification codes:} C12, C36, C55.}

\section{Introduction}
\label{section-introduction}

Recent empirical applications of instrumental variables (IV) estimation often involve many instruments that together may or may not be strongly relevant. For example, in a prominent paper by Angrist and Krueger (1991) that started the weak IV literature, the authors construct 180 instruments by interacting dummies for  the quarter of birth with state and year of birth, and use these instruments to study the effect of schooling on wage. Other examples include papers that employ an empirical strategy known as ``judge design'' (Maestas et al., 2013; Sampat and Williams, 2015; Dobbie et al., 2018). Fueled by rich administrative data, these papers use the exogenous assignment of cases  to judges as instruments for treatment. Since each judge can only process a certain number of cases out of the total court cases, the number of judges (the number of instruments) is usually proportional to the sample size.  Another example is the famous Fama-MacBeth procedure in Asset Pricing (Fama and MacBeth, 1973; Shanken, 1992), which is equivalent to  IV estimation procedure with the number of instruments proportional to the number of assets.

This paper answers three questions in an environment with many instruments: how to define weak identification, what to do if identification is weak, and how to pre-test for weak instruments.  We model many-instrument asymptotics by allowing the number of instruments to grow at most proportionally with the sample size. Firstly, we define weak identification for linear IV models with many instruments  by providing necessary and sufficient conditions for the existence of a consistent test.  Secondly,  we introduce a test that works when there are many instruments, but is also robust to weak identification and heteroscedasticity.  Finally, we propose a pre-test for weak identification. This pre-test forms the basis for a two-step procedure that is analogous to that of Stock and Yogo (2005). The two-step test controls size distortion under many-instrument asymptotics,  regardless of the strength of identification or the presence of heteroscedasticity.

We define weak identification as a situation where an analog of the concentration parameter divided by the \emph{square root of the number of instruments} stays bounded in large samples. We prove that even in a homoscedastic model with known covariance, an asymptotically consistent test does not exist if the ratio of the concentration parameter over the square root of the number of instruments stays bounded in large samples. Thus, a necessary condition for a consistent test to exist is that the concentration parameter grows faster than the square root of the number of instruments. Later, we show that this is also a sufficient condition by constructing a robust test that becomes consistent when this condition is satisfied.

We propose a new jackknifed  version of  the Anderson-Rubin (AR) test which is robust to both weak identification and heteroscedasticity in a model with  many instruments.  The new test uses an asymptotic approximation based on a Central Limit Theorem (CLT) for quadratic forms.  The new AR test has the correct size regardless of identification strength and becomes consistent as soon as the concentration parameter grows faster than the square root of the number of instruments.

As an important technical contribution, we introduce a novel variance estimator for the quadratic form CLT in the absence of a consistent estimator for the structural parameter. The target variance  is a quadratic form of the individual (heteroscedastic) variances of errors. We apply cross-fitting (Newey and Robins, 2018; Kline et al., 2020) to produce unbiased proxies for the individual variances of  errors. We adjust the quadratic form  to remove the bias due to correlations between proxies. We prove the consistency of the new estimator under the null and local alternatives under a wide range of identification scenarios.

Finally, we propose a new pre-test for weak identification which is easy to use and is consistent with our definition of weak identification. An empirical researcher can use our pre-test to decide between employing our jackknife AR test if the pre-test suggests that the identification is weak or a Wald test based on the Jackknife Instrumental Variable Estimator (JIVE, Angrist et al., 1999) if the pre-test suggests that the identification is strong. We guarantee the size of this two-step procedure.  Chao et al. (2012) prove that JIVE is consistent in a heteroscedastic model when the concentration parameter grows faster than the square root of the number of instruments. Chao et al. (2012) also derive a consistent estimator of the JIVE standard error. The  two-step procedure is appealing because
when identification is strong, the JIVE-Wald is more efficient and  easy to implement and report.

Our pre-test  is in the spirit of Stock and Yogo (2005), but it differs from theirs  in two important ways. Firstly, our pre-test allows for a general form of heteroscedasticity, while the pre-test proposed in Stock and Yogo (2005) works only under conditionally homoscedastic errors. Secondly, the Stock and Yogo (2005)   pre-test is designed for a  small number of instruments and is based on the Two-Stage Least Squares (TSLS) estimator. With many instruments TSLS is consistent only when the concentration parameter grows faster than the number of instruments, which makes the Stock and Yogo (2005) pre-test not very informative.

We apply our pre-test  to   Angrist and Krueger (1991) and find that their identification is strong.  Consequently the JIVE confidence set is reliable (has coverage within 5\% tolerance level of the declared coverage). Our weak identification-robust jackknife AR confidence set is somewhat wider than the JIVE confidence set but is still informative.

{\bf{Relation to the Literature}}. Our paper contributes  to both the literature on weak IV  and the literature on many instruments.  The weak IV literature relates identification strength to the size of the concentration parameter and proposes robust tests that work only when there are a small number of instruments. Generalizations to many weak instruments either strongly restrict the number of instruments (Andrews and Stock, 2007) or work only under homoscedasticity  (Anatolyev and Gospodinov, 2011). Crudu et al. (2020) recently proposed a jackknife-type AR test that is robust towards weak identification and heteroscedasticity. While employing a similar statistic, their test uses a variance estimator different from ours, that can be shown to lead to a power loss at distant alternatives and inconsistency of the test in some settings where our test is consistent.

The many weak instruments literature started with a prominent paper by Bekker (1994). It mostly establishes conditions for consistency and asymptotic gaussianity  for particular estimators.  For example,  Chao and Swanson (2005) show that in a homoscedastic model limited information maximum likelihood (LIML) and bias-corrected TSLS (\mbox{BTSLS}) are consistent when the concentration parameter grows faster than the square root of the number of instruments. In a heteroscedastic model, consistency of LIML and \mbox{BTSLS} requires that the concentration parameter grows faster than the number of instruments. By contrast, JIVE remains consistent when the concentration parameter grows faster than the square root of the number of instruments (Chao et al., 2012). Our paper shows that the condition in Chao et al. (2012) is necessary for consistency and if it is violated it is impossible to consistently distinguish between any two values of the structural parameter.

The remainder of this paper is organized as follows. Section \ref{section - new} summarizes our proposal for empirical researchers. In Section \ref{section-weak id} we introduce our definition of weak identification in an environment with many instruments. In Section \ref{section- AR} we construct the jackknife AR test and establish its power properties. In Section \ref{section - pretest} we present the pre-test and prove that it controls size.   Section \ref{section - simulation} conducts a simulation exercise inspired by Angrist and Frandsen (2019), and Section \ref{section- conclusion} concludes. Some proofs and additional results may be found in the Supplementary Appendix.

\section{Many Weak Instruments: Empirical Practice}\label{section - new}
In empirical applications  using instrumental variables, concerns about weak identification are widespread. The  current consensus practice is to report the first stage $F$ statistic and as long as it is above 10, researchers are allowed to rely on standard $t$-statistics inferences. This practice has foundations in Stock and Yogo (2005) which showed that the concentration parameter fully characterizes the  size distortion of the TSLS-Wald test,  and empirically the concentration parameter can be judged based on the first stage $F$ statistics. This result has been obtained under the assumptions of homoscedasticity and for a fixed number of instruments.

While the first stage $F$ pre-test  provides reasonable classification for homoscedastic IV models with a small number of instruments, it is inadequate for settings with many instruments. Hansen et al (2008)  argue  that the TSLS estimator should not be used in applications with many instruments as it becomes very biased. They also  argue that  a low first stage $F$ statistic is not always indicative of a weak identification issue and $t$-statistics inferences based on  more appropriate estimators other than TSLS, along with corrected standard errors,  may still be reliable. Estimators with known good properties in heteroscedastic settings with many instruments include JIVE (Chao and Swanson, 2005) and heteroscedasticity-robust Fuller (Hausman et al, 2012).

While theoretical econometrics literature provides recommendations on the choice of estimator, it is largely silent on how to determine whether the concerns of weak identification are valid in a given data set with many instruments. This prompted  empirical researchers to formulate econometric arguments and perform simulation studies to support their usage of $t$-statistics. For example,  Bhuller et al (2020), recently published in the \emph{Journal of Political Economy}, used judges design instruments to study the effects of incarceration on recidivism and employment. Concerned about potentially having many weak instruments, Bhuller et al (2020) included a 10-page-long Appendix D with a simulation study to support their usage of the JIVE $t$-statistic.

Our paper proposes a new recipe for empirical researchers to gauge weak identification in applications with many instruments. Specifically, we argue that theoretically, the strength of identification is measured by the concentration parameter \emph{divided by the square root of the number of instruments}. This is in contrast to the first stage $F$ statistics from Stock and Yogo (2005), which implicitly divide the concentration parameter  by the number of instruments. We suggest  applied researchers  calculate a new pre-test $\widetilde{F}$ (see equation (\ref{eq: F statistics})) and compare it to a cutoff of 4.14. If $\widetilde{F}$ is above the cutoff then the researcher can rely on the JIVE $t$-statistic  with the caveats analogous to Stock and Yogo (2005): Namely, the size distortions of the JIVE $t$-statistic are within 5\% tolerance level of the nominal size. If $\widetilde{F}$ is below the cutoff we suggest researchers report a confidence set obtained  by inversion of our newly proposed weak-identification robust jackknife AR test (see Equation (\ref{eq: def of AR})). As discussed in Section~\ref{section - pretest}, applied researchers may also choose other cutoffs depending on their tolerance level of size distortions. Here we illustrate this recipe with an example from Angrist and Krueger (1991) (hereafter referred to as AK91).

AK91 provided a motivating example for the weak identification literature,  starting with the seminal work by Bound et al. (1995).   Staiger and Stock (1997) suggested that the relatively low value of the first stage $F$ statistic  can be seen as a sign of potentially weak instruments in the AK91  application. Hansen et al. (2008) argued that \emph{many instruments} may be a more relevant description of the identification issue encountered in AK91.  They suggested that estimators other than the TSLS may restore the reliability of standard inferences. We resolve the controversy of whether the instruments are weak in this example utilizing a formal pre-test.

The original AK91 application estimated the effect of schooling ($X_{i}$) on log weekly wage ($Y_{i}$) using quarter of birth as instruments in a sample from the 1980 census of 329,509 men born in 1930-39. There are multiple specifications in the original AK91 study.  We focus on the specification with 180 instruments and also  on an extension of this specification using 1,530 instruments.  The 180 instruments include  30 quarter and year of birth interactions (QOB-YOB) and 150 quarter and state of birth interactions (QOB-POB).  For the second specification with 1,530 instruments, we also include full interactions among QOB-YOB-POB.  Table~\ref{tab:AK91} reports the first stage $F$ statistics (FF), our proposed pre-test statistics $\widetilde{F}$,  5\% and 2\% confidence sets based on the JIVE $t$-statistic and the jackknife AR statistic proposed in this paper.
  
 While the first stage $F$ statistic is below 10 and the current empirical practice would point towards weak identification for both specifications, the instruments turn out to be strong in both specifications based on our pre-test.  According to  the results of this paper discussed in Section~\ref{section - pretest}, both of the reported confidence sets based on a nominal 5\% JIVE $t$-test are reliable with the same caveats as in Stock and Yogo (2005), namely, the actual rejection rate under the null hypothesis does not exceed 10\%.  This pre-test is based on the statistic $\widetilde{F}$ and rejects whenever $\widetilde{F} > 4.14$.  Based on the pre-test, the empirical researcher may report the JIVE confidence set only, and not the identification-robust AR confidence set. 
 
 This two-step procedure is similar to that popularized by Stock and Yogo's (2005).  Choosing between JIVE and AR confidence set to report based on the pretest in the first step guarantees that the reported confident set from this two-step procedure has an overall size of 15\%.  If the applied researcher prefers that the two-step procedure has an overall size of 5\%, results in Section~\ref{section - pretest} suggest using a higher cutoff of 9.98 for $\widetilde{F}$ and smaller nominal sizes to construct the confidence sets. Specifically, the applied researcher should choose between a nominal 98\%-level JIVE confidence set and a nominal 98\%-level AR confidence set.    In this case, for the specification with 180 instruments, the applied researcher can still report the  JIVE confidence set as the corresponding $\widetilde{F}$ is greater than 9.98.  However, for the specification with 1530 instruments, the applied researcher needs to report the identification-robust AR confidence set instead as the corresponding $\widetilde{F}$ is less than 9.98.
 
 An alternative to pre-test is to always report a robust confidence set, which would be the 5\% jackknife AR confidence set in this case. We do see that the AR confidence sets are wider, yet still informative.

\begin{table}

\begin{centering}
\small{\begin{tabular}{ccccccc}
\hline
 &FF &$\widetilde{F}$ & JIVE-$t$  & Jackknife AR   & JIVE-$t$  & Jackknife AR    \\
 & & &(5\%) &   (5\%) &(2\%) &   (2\%) \tabularnewline
\hline
\hline
180 instruments & 2.43 &13.42 & \textbf{{[}0.066,0.13{]} }& {[}0.008,0.20{]} & \emph{{[}0.059,0.14{]}} & {[}0.0003, 0.21{]}\tabularnewline
\hline
1530 instruments  & 1.27 & 6.17 &\textbf{ {[}0.024,0.12{]}} & {[}-0.047, 0.20{]}& {[}0.015,0.13{]} & \emph{{[} -0.066,0.22{]}}\tabularnewline
\hline
\end{tabular}}
\par\end{centering}
\caption{\label{tab:AK91} AK91 Pre-test Results}

{\footnotesize{\emph{Notes: }Results on pre-tests for weak identification and confidence sets for the IV  specification underlying Table VII
Column (6) of Angrist and Krueger (1991) using the original data. FF is the first stage F statistic of  Stock and Yogo (2005), $\widetilde{F}$ is the statistic  introduced in (\ref{eq: F statistics}).    The  jackknife AR confidence set is based on analytical test inversion. The confidence sets reported by the two-step procedure with Stock and Yogo's guarantee are in bold. The confidence sets reported by the two-step procedure with overall size of 5\% is in italic.}}

\end{table}

\section{Weak Identification with Many Instruments}\label{section-weak id}
We study the linear IV regression with a scalar outcome $Y_i$, a potentially endogenous scalar regressor $X_i$ and a $K\times 1$ vector of instrumental variables  $Z_i$:
\begin{equation}\label{eq: iv model}
\left\{ \begin{array}{c}
          Y_i=\beta X_i+e_i ,\\
          X_i=\Pi_i+v_i,
        \end{array}
\right.
\end{equation}
for $ i=1,..., N.$ We denote $Y$ to be the $N\times 1$ vector of outcome and $X$ to be the $N\times 1$ vector of endogenous regressors. We collect the transpose of $Z_i$ in each row of $Z$, a $N\times K$ matrix of instruments.  We denote $\Pi_i=\mathbb{E}[X_i|Z_i]$ and allow the instruments to affect the endogenous regressor in a non-linear way. All results in this paper hold conditionally on a realization of the instruments.
Thus, we treat  the instruments as fixed (non-random) and $\Pi_i$ as some constants.  We collect $\Pi_i$ in   $\Pi$, a $N\times 1$ vector.
The mean-zero errors $(e_i,v_i)$ are independent across $i$ but not identically distributed and may be heteroscedastic.
We assume without loss of generality that there are no controls included in our model as they may be partialled out.

Weak identification under small $K$ is studied extensively in the weak IV literature.  For Gaussian homoscedastic errors $(e_i,v_i)$ and linear first stage ($\Pi_i=\pi'Z_i$), the strength of the instruments corresponds directly to  the concentration parameter, $\frac{\pi^\prime Z'Z\pi}{\sigma_v^2}$ where $\sigma_v^2=Var(v_i)$. The concentration parameter equals  the signal-to-noise ratio in the first-stage regression and is related to the bias of the TSLS estimator and the quality of Gaussian approximation for the  TSLS $t$-statistic.  For the general case with homoscedastic errors,  Staiger and Stock (1997) introduced weak instrument-asymptotics in which one considers a sequence of models so that the concentration parameter converges to a constant as $N\to\infty$.  Under this asymptotic embedding, neither a consistent estimator of $\beta$ nor a consistent test of the null hypothesis that $\beta$ equals some scalar exists, and the test based on the TSLS $t$-statistic severely over-rejects.

The magnitude of the concentration parameter  is not a good indicator of identification strength when the number of instruments is large.   Inspired by Bekker (1994), we model large $K$ by considering $K\to\infty$ as $N\to\infty$, with the only restriction  that $K$ is at most a fraction of $N$. Under this many instrument-asymptotics, Theorem \ref{thm: negative result}  below shows that the re-scaled concentration parameter $\frac{\pi^\prime Z'Z\pi}{\sigma_v^2\sqrt{K}}$ provides a  characterization of weak identification in terms of the consistency of tests.

\begin{theorem}\label{thm: negative result}
Assume we have a sample from model (\ref{eq: iv model}) with linear first stage $\Pi_i=\pi'Z_i$. Consider the reduced-form errors $(u_i,v_i)$ where $u_i = Y_i-\beta\pi'Z_i$. Assume the reduced-form errors are independently drawn from a Gaussian distribution $\mathcal{N}(0,\Omega)$ with a known nonsingular covariance matrix $\Omega$. Assume that  the  $K\times K$ matrix $Z'Z$  has rank $K$ and  $K\to\infty$ as $N\to\infty$. For any sample of size $N$ let $\Psi_N$ be the class of all  tests of size $\alpha$ for testing  the hypothesis $H_0:\beta=\beta_0$, that is, any $\psi\in \Psi_N$ is a measurable function from $\{(Y_i,X_i,Z_i), i=1,..., N\}$ to the interval $[0,1]$ such that $\mathbb{E}_{\beta_0,\pi}\psi\leq \alpha$ for any value of $\pi\in \mathbb{R}^K$. Then for any $\beta^*\neq \beta_0$ we have
$$
\lim\sup_{N\to\infty}\max_{\psi\in\Psi_N}\left(\min_{\pi: \frac{\pi'Z'Z\pi}{\sigma_v^2\sqrt{K}}= C}\mathbb{E}_{\beta^*,\pi}\psi\right)<1.
$$
\end{theorem}

The setting considered in Theorem \ref{thm: negative result} is quite favorable: the first stage is linear, errors are Gaussian and homoscedastic with known covariance matrix. So the only unknown  parameters are $\beta$ and $\pi$. Theorem \ref{thm: negative result}  states that even in this favorable setting there exists no test that consistently differentiates any $\beta^*$ from $\beta_0$  if the ratio of the concentration parameter to the square root of the number of instruments is bounded. Indeed, for any test $\psi$ we can find its  guaranteed power  $\mathbb{E}_{\beta^*,\pi}\psi$ by minimizing over the alternatives $(\beta^*,\pi)$ with bounded ratio of the concentration parameter over $\sqrt{K}$. We show that even in this favorable  setting the  test that achieves  the maximum guaranteed power  has guaranteed power strictly less than one asymptotically. With heteroscedasticity of  unknown form, sufficient statistics of low dimensions are not known, making the setting even less favorable. Later we show that in a more general heteroscedastic model we can construct a robust test that becomes consistent when $\frac{\Pi^\prime \Pi}{\sqrt{K}}\to\infty$.

Theorem \ref{thm: negative result} can also be used to characterize weak identification in terms of consistent estimation since it implies there exists no consistent estimator for $\beta$ when  the ratio of the concentration parameter to $\sqrt{K}$ is bounded.  Our result complements the literature on estimation with many instruments. Chao and Swanson (2005) show that with homoscedastic errors, when $K$  grows proportionally to the sample size the TSLS estimator is consistent only if the concentration parameter grows faster than the number of instruments $K$, while LIML and \mbox{BTSLS} estimators are consistent when the concentration parameter grows faster than $\sqrt{K}$.  However, under heteroscedasticity, even when $\frac{\pi^\prime Z'Z\pi}{\sqrt{K}}\to\infty$,  LIML and \mbox{BTSLS} become inconsistent, but JIVE is still consistent, according to Chao et al. (2012).

The proof of Theorem \ref{thm: negative result} builds on several classical papers.  Following the approach of Andrews et al. (2006), we first reduce the class of tests to those  based on a sufficient statistic. Among these tests, the minimal power is achieved by a test invariant to rotations of the instruments.  This observation allows us to further reduce our attention to invariant tests, which depend on the data only through its maximal invariant under rotations.  Then we derive a limit experiment for $K\to\infty$ similar to that derived in Andrews and Stock (2007).  In this limit experiment the minimax power is less than one.  Finally we use the  argument of M{\"u}eller (2011) to bound the desired asymptotic minimax power using the minimax power obtained in the limit experiment.

\section{Jackknife AR}\label{section- AR}

The goal of this section is to introduce a test robust to weak identification  in the heteroscedastic IV model when the number of instruments, $K$, is large.

The existing weak IV literature proposes several weak identification-robust tests of the null hypothesis $H_0:\beta=\beta_0$, when $K$ is small.  These tests have correct size when the identification is weak and become consistent when the identification is strong.
One example is the AR test.
Specifically, the IV model (\ref{eq: iv model})  implies that  under a given null hypothesis $H_0:\beta=\beta_0$, the exogeneity assumption holds $\mathbb{E}[Z'e(\beta_0)]=0$ for the implied error $e(\beta_0)=Y-\beta_0X$.
Then under mild assumptions, the scaled sample analog $\frac{1}{\sqrt{N}}Z'e(\beta_0)\Rightarrow N(0,\Sigma)$ satisfies a $K$-dimensional CLT. The AR statistic is defined as
$
\frac{1}{N}e(\beta_0)'Z\widehat\Sigma^{-1}Z'e(\beta_0)
$, where $\widehat\Sigma$ is a consistent estimator of $Var\left(\frac{1}{\sqrt{N}}Z'e\right)$.  The AR test rejects the null hypothesis when the AR statistic exceeds the $(1-\alpha)$ quantile of the $\chi^2_K$ distribution. The AR test has asymptotically correct size regardless of the value of the first stage coefficients $\Pi_i$ and is asymptotically consistent when an analog of the concentration parameter grows to infinity.

Generalizing  the AR statistic to the large-$K$ setting is challenging for multiple reasons.  Firstly, the covariance matrix $\Sigma$ has dimension $K\times K$.  Its consistent estimation is problematic if not impossible under general heteroscedasticity.  Secondly, the AR statistic under the null has an improperly centered limit distribution because $\chi_K^2$ has a very large mean. Thirdly, the $K$-dimensional CLT provides a poor approximation to the AR statistic when $K$ is large.

We propose an analog of the AR test that is heteroscedasticity-robust and weak identification-robust in the presence of  a large number of instruments. Denote the projection matrix $P=Z(Z'Z)^{-1}Z'$. Our test rejects the null of $H_0:\beta=\beta_0$ when the jackknife AR statistic
\begin{equation}\label{eq: def of AR}
AR(\beta_0)=\frac{1}{\sqrt{K}\sqrt{\widehat\Phi}}\sum_{i=1}^N\sum_{j\neq i}P_{ij}e_i(\beta_0)e_j(\beta_0)
\end{equation}
exceeds the $(1-\alpha)$ quantile of the standard normal distribution. We defer the discussion of the estimator of the variance $\widehat\Phi$ to the next subsection.

To address the challenges with the existing AR statistic, the AR statistic we propose uses the default homoscedasticity-inspired weighting $(Z'Z)^{-1}$ in place of $\widehat\Sigma^{-1}$.  With the $(Z'Z)^{-1}$ weighting, the existing AR statistic has a quadratic form $e(\beta_0)'Pe(\beta_0)$.  However, this quadratic form is not centered at zero as it contains the term $\sum_{i=1}^NP_{ii}e_i^2$, and each  summand has positive mean.  We thus remove this term from the quadratic form.  This  re-centering can be referred to as leave-one-out or  jackknife. In the context of consistent estimation under many instruments, this leave-one-out idea was introduced by Angrist et al. (1999) and fruitfully exploited in a number of papers including Hausman et al. (2012) and Chao et al. (2012). Recently, this idea has been used in Chao et
al. (2014) and  Crudu et al. (2020). In order to create a test of correct size based on our AR statistic, we use a CLT for quadratic forms proved in Chao et al. (2012) that is restated below.

\begin{ass}\label{ass: projection matrix}
Assume $P$ is an $N\times N$ projection matrix of rank $K$, $K\to\infty$ as $N\to\infty$ and there exists a constant $\delta$ such that $P_{ii}\leq \delta<1$.
\end{ass}

\begin{lemma}\label{lemma: CLT} (Chao et al., 2012)
Let Assumption \ref{ass: projection matrix} hold for matrix $P$.  Assume the errors $\eta_i$ are independent, $\mathbb{E}\eta_i=0$, and there exists a constant $C$ such that $\max_i\mathbb{E}\eta_i^4<C$, then
$$
\frac{1}{\sqrt{K}\sqrt{\Phi}}\sum_{i=1}^N\sum_{j\neq i}P_{ij}\eta_i\eta_j\Rightarrow \mathcal{N}(0,1),
$$
where $\Phi=\frac{2}{K}\sum_{i=1}^N\sum_{j\neq i}P_{ij}^2Var(\eta_i)Var(\eta_j)$.
\end{lemma}

The assumption $P_{ii}\leq \delta<1$ implies that $\frac{K}{N}=\frac{1}{N}\sum_{i=1}^NP_{ii}\leq \delta<1$. This assumption is often referred to as a balanced design assumption. In the case of group-dummies instruments, $P_{ii}$ is equal to the ratio of the size of the group that observation $i$ belongs to over $N$. Assumption \ref{ass: projection matrix} can be checked for any specific design.

While Lemma \ref{lemma: CLT} requires $K\to\infty$, the Gaussian approximation may work well for smaller $K$ as well. For example, if $K$ is fixed and errors are homoscedastic,  then
$$
\frac{1}{\sqrt{K}\sqrt{\Phi}}\sum_{i=1}^N\sum_{j\neq i}P_{ij}\eta_i\eta_j\Rightarrow \frac{\chi^2_K-K}{\sqrt{2K}} ~~~\mbox{     as    }~~~ N\to\infty.
$$
We prove this statement in the Supplementary Appendix S4.
While the limit here is not Gaussian it is very well approximated by a standard normal distribution even for relatively small $K$. The random variable $\frac{\chi^2_K-K}{\sqrt{2K}}$ exceeds  the  95\% quantile of the standard normal distribution at most 7\% of the time for all $K$, and  at most 6\% of the time for $K>40$.

\subsection{Variance estimation}\label{subsection- variance}

In order to conduct asymptotically valid inference based on the normal approximation in Lemma \ref{lemma: CLT}, we need an estimator for the scale parameter $\Phi$, which is consistent under the null.
One `naive' estimator that achieves this is
$
\widehat\Phi_1=\frac{2}{K}\sum_{i=1}^N\sum_{j\neq i}P_{ij}^2e_i^2(\beta_0)e_j^2(\beta_0),
$
which uses the square of the implied error as an estimator for the $i$-th error variance. Under the null when $e_i(\beta_0)=e_i$, the estimator $\widehat\Phi_1$ is consistent under  relatively mild conditions.  However, using $\widehat\Phi_1$ in  a test would result  in poor power. To see this, note that under an alternative value of the parameter  $\beta=\beta_0+\Delta$,  we can plug in the first stage and write the implied error $e_i(\beta_0)=Y_i-\beta_0X_i$ as the sum of a non-trivial mean $\Delta  \Pi_i$ and a mean-zero random term $\eta_i = e_i+\Delta v_i$:
\begin{align}\label{eq: error vs implied}
e_i(\beta_0)= \Delta \Pi_i + \eta_i.
\end{align}
The AR statistics of a form similar to (\ref{eq: def of AR}) with $\widehat\Phi_1$ has been recently and independently proposed by Crudu et al. (2020). The aforementioned paper establishes  robustness  of their proposed test towards weak identification and heteroscedasticity in terms of size.  While squaring $e_i(\beta_0)$ makes it an unbiased estimator for $Var(e_i)$ under the null, it is biased under the alternative when $\Delta \neq 0$. The bias in $\widehat\Phi_1$ grows at the same order as the fourth power of $\Delta$, which brings down the power of the test against distant alternatives.  In Section~\ref{subsection - power}, we discuss the power implications of the `naive' estimator in more detail.

In order to remove the bias in $e_i^2(\beta_0)$ under the alternatives, one may residualize the implied error before squaring. However, this introduces a bias under the null. Denote $M=I-P$ and let $M_i$ be the $i$th row of $M$.  Even under the null, the squared residualized error is biased $\mathbb{E}(M_ie)^2\neq Var(e_i)$. This is because the  squared residual contains not only the squared error $e_i$ but also the square of regression estimation mistake. The latter can be large when the number of regressors $K$ is large.

This bias can be removed successfully using the cross-fit variance estimator suggested in Kline et al. (2020) and Newey and Robins (2018). Namely, they show that a product of the implied error and residual achieves both goals: it removes the linearly predictable part of the implied error and  remains an unbiased estimator of the variance
$$
\mathbb{E}\left[\frac{e_iM_ie}{M_{ii}}\right]=Var(e_i).
$$

Our challenge  is that the scale parameter $\Phi$ defined in Lemma \ref{lemma: CLT} is a quadratic form with a double summation. Residuals $M_ie(\beta_0)$ and $M_je(\beta_0)$ are correlated since they contain the same estimation mistake. One can show that
$$
\mathbb{E}\left[e_iM_iee_jM_je\right]=(M_{ii}M_{jj}+M_{ij}^2)Var(e_i)Var(e_j).
$$
Our proposed estimator of the scale parameter $\Phi$ re-weights each term in the summation to remove the bias described above:
\begin{equation}\label{eq: def of hat Phi}
\widehat\Phi=\frac{2}{K}\sum_{i=1}^N\sum_{j\neq i}\frac{P_{ij}^2}{M_{ii}M_{jj}+M_{ij}^2}\left[e_i(\beta_0)M_ie(\beta_0)\right]\left[e_j(\beta_0)M_je(\beta_0)\right].
\end{equation}
We establish the consistency of $\widehat\Phi$ under the null and  extend this result  to local alternatives.

\begin{ass}\label{ass: errors}
Errors  $\epsilon_i, i=1,...,N$  are independent with $\mathbb{E}\epsilon_i=0$,  $\max_i\mathbb{E}\|\epsilon_i\|^6<\infty$, and for some constants $c^*$ and $C^*$ that do not depend on $N $
$$c^*\leq \min_i \min_{x}\frac{x'Var(\epsilon_i)x}{x'x}\leq \max_i\max_x\frac{x'Var(\epsilon_i)x}{x'x}\leq C^*.$$
\end{ass}

\begin{theorem}\label{thm: consistency of variance null}
Let Assumption \ref{ass: projection matrix} hold for matrix $P$ and Assumption \ref{ass: errors} hold for  errors $e_i$, then for $\beta=\beta_0$, we have $
  \frac{\widehat\Phi}{\Phi}\to^p 1$ as    $ N\to\infty.
$
\end{theorem}
Theorem \ref{thm: consistency of variance null} combined with Lemma \ref{lemma: CLT} implies that under the null $H_0:\beta=\beta_0$ our proposed AR statistic has an asymptotically standard normal distribution.  Since no assumption about identification is made, the resulting AR test has asymptotically correct size regardless of the  strength of identification.

\begin{theorem}\label{thm: consistency of variance alternative}
Let Assumption \ref{ass: projection matrix} hold for matrix $P$  and  Assumption \ref{ass: errors} hold for  errors $\epsilon_i=(e_i,v_i)'$, and $\Pi'M\Pi\leq\frac{C}{K}\Pi'\Pi$.  Then for  $\beta=\beta_0+\Delta$, where $\Delta$ may depend on $N$  such that $\Delta^2\cdot\frac{\Pi'\Pi}{K}\to 0$, we have $  \frac{\widehat\Phi}{\Phi}\to^p 1$ as    $ N\to\infty.$
\end{theorem}

Theorem \ref{thm: consistency of variance alternative}  establishes the consistency of the variance estimator when the null hypothesis does not hold. We use Theorem \ref{thm: consistency of variance alternative} to derive local power curves of the AR test discussed in the next section. The variance estimator (\ref{eq: def of hat Phi}) residualizes the implied errors $M_ie(\beta_0)$  to remove non-trivial mean of $e(\beta_0)$ under the alternative. The residualization is complete if the first stage is linear $\Pi_i=\pi'Z_i$. We do not impose such an assumption in Theorem \ref{thm: consistency of variance alternative}.  Instead we require that the approximation of $\Pi_i$  by a linear combination of instruments improves with the number of instruments as measured by the  norm of the approximation mistake, $\Pi'M\Pi$. In their Assumption 4, Chao et al. (2012) impose that $\frac{\Pi'M\Pi}{N}\to 0$, which may be weaker or stronger than our assumption $\Pi'M\Pi\leq\frac{C}{K}\Pi'\Pi$ depending on the  identification strength. The variance estimation in Chao et al. (2012) is valid only under strong identification as it relies on the consistency of the JIVE estimator. The residuals from structural equation, with the JIVE  estimate for $\beta$ plugged in, approximate the structural errors well.  In contrast, our variance estimator remains valid under weak identification when no consistent estimator for $\beta$ exists.  This is why we need stricter assumptions on the linear approximation to produce reliable residuals under weak identification.

\subsection{Power of the Jackknife AR test}\label{subsection - power}
Let us introduce a jackknife measure of the information contained in the instruments:
$$
\mu^2=\sum_{i=1}^N\sum_{j\neq i}P_{ij}\Pi_i\Pi_j.
$$
For the linear first stage $\Pi_i=\pi'Z_i$, we have $\mu^2=\pi'Z'Z\pi-\sum_{i=1}^NP_{ii}(\pi'Z_i)^2$.  Assumption \ref{ass: projection matrix} guarantees that $(1-\delta)\pi'Z'Z\pi\leq\mu^2\leq \pi'Z'Z\pi.$ Thus, the two measures $\frac{\mu^2}{\sqrt{K}}$ and $\frac{\pi'Z'Z\pi}{\sqrt{K}}$ are of the same order and increase to infinity or not simultaneously. In the general case where the instruments may affect the endogenous regressor in an arbitrarily non-linear way, the linear IV regression only uses the projection of $\Pi$ onto the linear space of the instruments.  Thus the projection matrix appears naturally in our measure of identification strength.  The parameter $\mu^2$ can be considered as a jackknife generalization of the parameter $\pi'Z'Z\pi$ to non-linear case.
\begin{theorem}\label{thm:AR power1}
Let $\mathbb{P}_{\beta}$ be a probability measure describing the distribution of $AR(\beta_0)$ defined in (\ref{eq: def of AR}) and (\ref{eq: def of hat Phi}) under model (\ref{eq: iv model}) with  parameter $\beta=\beta_0+\Delta$.
Assume that the sequence of  first stage parameters $\Pi$ satisfies the following assumptions: $\Pi'M\Pi\leq\frac{C}{K}\Pi'\Pi$  and $\frac{\Pi'\Pi}{K}\to 0$ as $N\to \infty$.
If Assumption \ref{ass: projection matrix}  holds and the errors $\epsilon_i=(e_i,v_i)'$ satisfy Assumption \ref{ass: errors}, then for any  positive constant $c$ we have:
\begin{equation}\label{eq: them about local power}
\lim_{N\to\infty}\sup_{ |\Delta|^2\leq c}\sup_z\left| \mathbb{P}_{\beta}\{AR(\beta_0)<z\}-F\left( z - \frac{\Delta^2\mu^2}{\sqrt{K\Phi}}\right)\right|=0,
\end{equation}
where  $F(\cdot)$ is the standard normal cdf.
If the sequence of first stage parameters additionally satisfies the condition $\frac{\mu^2}{\sqrt{K\Phi}}\to\infty$, then  for any fixed $\Delta\neq 0$ the jackknife AR test is asymptotically consistent:
$$
\lim_{N\to\infty}\mathbb{P}_{\beta}\{AR(\beta_0) \geq z_{1-\alpha}\}=1
$$
where $z_{1-\alpha}$ is the $(1-\alpha)$ quantile of the standard normal distribution.
\end{theorem}
Equation (\ref{eq: them about local power}) of Theorem \ref{thm:AR power1}  characterizes the local power curves of the jackknife AR test.  The power under the alternative $\beta=\beta_0+\Delta$ is a function of  the distance $\Delta$ between the alternative $\beta$ and the null $\beta_0$, the number of instruments $K$, a measure of identification strength $\mu^2$ and the degree of uncertainty $\sqrt{\Phi}$.  Our jackknife AR statistic can be negative, unlike the AR statistic  from the small-$K$ case which is always non-negative. We reject the null when $AR(\beta_0)$ exceeds the $(1-\alpha)$ quantile of the standard normal distribution. Under the alternative $\beta=\beta_0+\Delta$, the AR statistics has a positive drift and produces non-trivial power for both positive and negative $\Delta$.
The second statement of Theorem \ref{thm:AR power1} shows that the AR test  consistently distinguishes $\beta$ from $\beta_0$ as long as $\frac{\mu^2}{\sqrt{K}\sqrt{\Phi}}\to\infty$.

Theorem \ref{thm:AR power1} implies that $\frac{\mu^2}{\sqrt{K}}\to\infty$ is a sufficient condition for the consistency  of the jackknife AR test in a model with a linear first stage .
This complements Theorem \ref{thm: negative result}  which implies  that  $\frac{\pi'Z'Z\pi}{\sqrt{K}}\to\infty$ is necessary for the consistency of any test. This condition has appeared before in Chao et al. (2012) as  a sufficient condition for the consistency of the JIVE estimator and asymptotic validity and consistency of the JIVE $t$-test. The important difference between the proposed jackknife AR test and the JIVE $t$-test is that even under weak identification ($\frac{\pi'Z'Z\pi}{\sqrt{K}}\not\to\infty$), the former maintains asymptotically valid size, while the latter does not.
It is worth noticing that the condition $\frac{\Pi'\Pi}{K}\to0$  imposed by Theorem \ref{thm:AR power1} is quite weak as it covers both weakly and strongly identified cases.

\paragraph{ Power implications of variance estimation.}
While the leave-one-out AR test with our proposed cross-fit variance estimator is consistent against fixed alternatives when identification is strong, the same test with a `naive' variance estimator $\widehat{\Phi}_1$ is in general not consistent. The difference between the implied error $e_i(\beta_0)$ and $\eta_i$ as defined in equation (\ref{eq: error vs implied}) results in that  the difference between $\widehat{\Phi}_1$ and $\Phi$  is a  fourth degree polynomial of $\Delta$. This makes the stochastic shift for the AR  statistic with the naive variance estimator to stabilize at the finite level when $\Delta\to\pm\infty$:
$$
\frac{\Delta^2\mu^2}{\sqrt{K}\sqrt{\widehat{\Phi}_1}}\approx \frac{\Delta^2\mu^2}{\sqrt{K}\sqrt{\Phi}}\sqrt{\frac{\Phi}{c\Delta^4+\Phi}}\to C_{\pm} \mbox{  as }\Delta\to\pm\infty,
$$
while it increases unboundedly for the statistic with the cross-fit variance estimator.
Here $c=\frac{2}{K}\sum_{i=1}^N\sum_{j\neq i}P_{ij}^2\Pi_i^2\Pi_j^2$.

Theoretical inconsistency of a test may or may not result in power differences of empirical relevance for commonly used significance levels. This depends partially on whether the level at which the stochastic shift stabilizes is above the typically used critical values.  In  very strongly identified cases where  $\frac{\mu^2}{\sqrt{K}}$ is large, we may detect no significant difference between two statistics for alternatives with relatively small $\Delta$, implying a small power difference.  While there is an increasingly large difference in realized values of statistics for large $\Delta$, such a difference might not translate to a power difference either since both tests would reject.   For example, in the AK91 example discussed in Section~\ref{section - new}, using the `naive' variance estimator $\widehat{\Phi}_1$ would yield nearly identical jackknife AR confidence sets.  For the specification that uses 1530 instruments, jackknife AR  confidence sets based on the `naive' variance estimator are $[-0.048, 0.202]$ (5\%) and $[-0.662, 0.224]$ (2\%), which are very close to the ones based on $\widehat\Phi$  as reported in Table~\ref{tab:AK91}.

\begin{figure}[t!]
     \centering

     \begin{subfigure}[t]{0.5\textwidth}
         \centering
   \includegraphics[width=\textwidth]{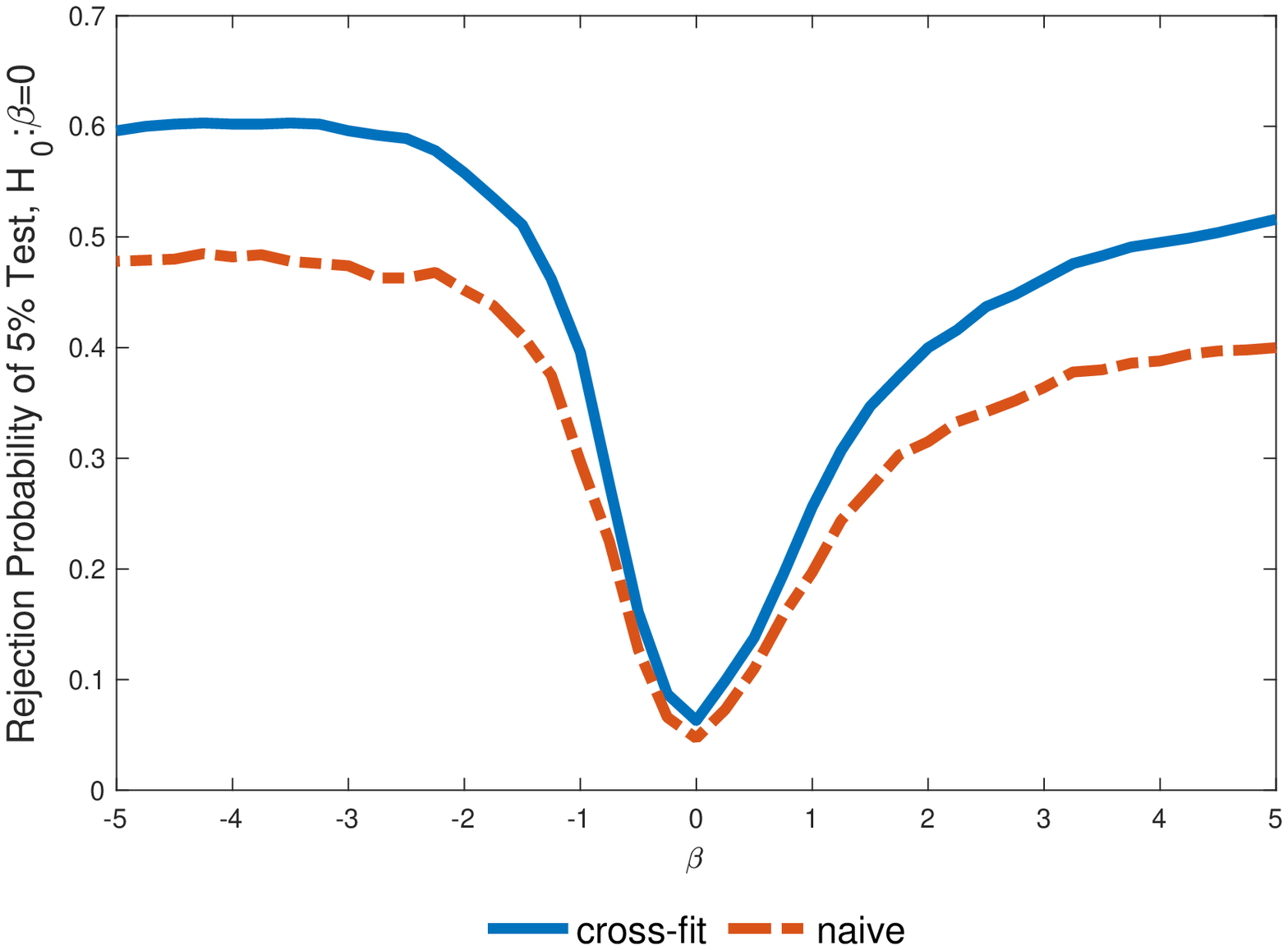}
         \caption{sparse,$\frac{\mu^2}{\sqrt{K}}=2.5$}
     \end{subfigure}%
    ~
     \begin{subfigure}[t]{0.5\textwidth}
         \centering
   \includegraphics[width=\textwidth]{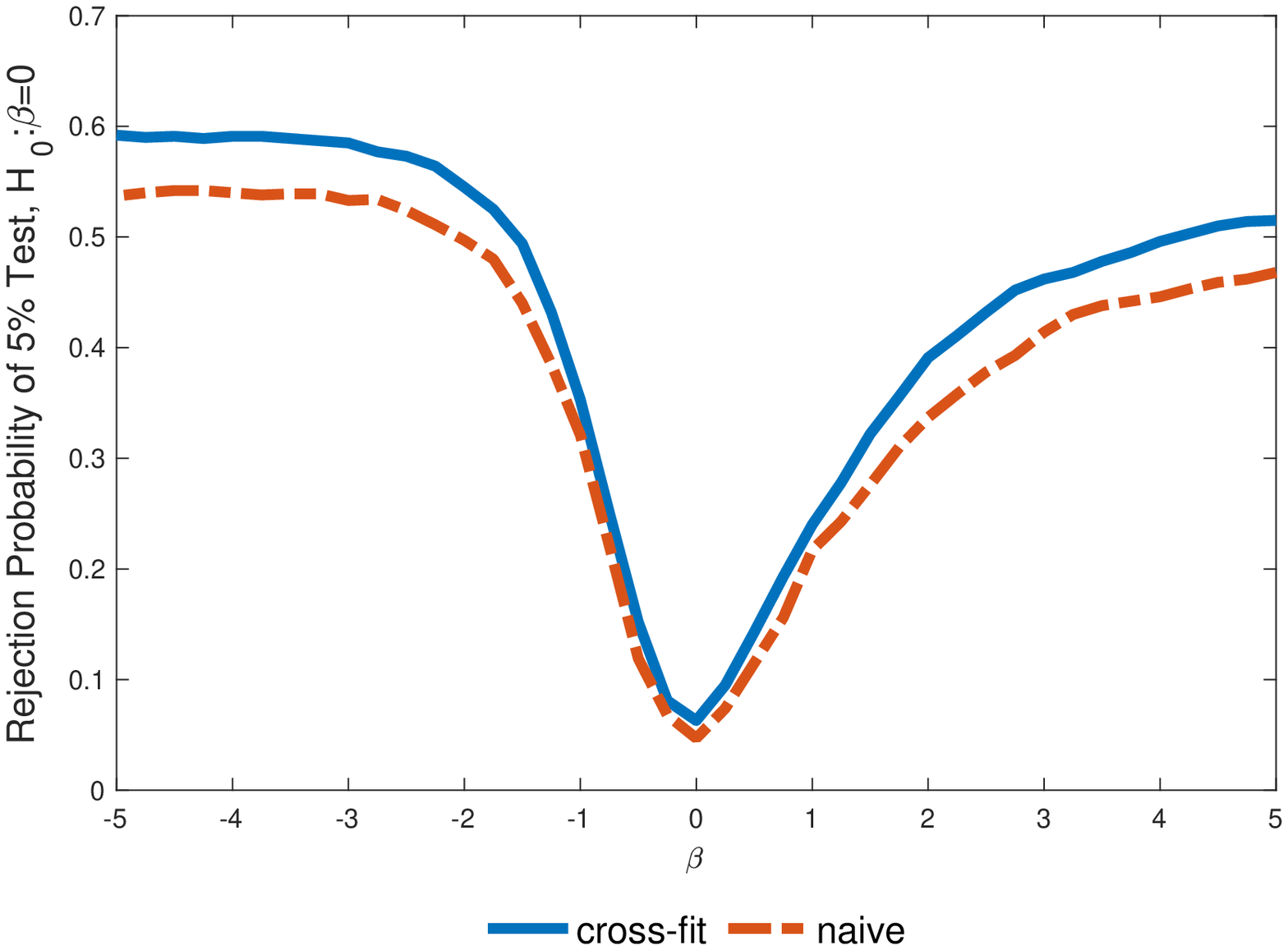}
         \caption{dense,$\frac{\mu^2}{\sqrt{K}}=2.5$}
     \end{subfigure}%
     
\caption{Power curves for leave-one-out AR tests with cross-fit (blue line) and naive (red dash) variance estimators under sparse vs. dense first stage.  The instruments are $K=40$ balanced group indicators. Sample size is $N=200$. Number of simulation draws is 1,000. Details of the simulation design can be found in the Appendix.}
        \label{fig:power dense}
\end{figure}

We find larger power differences for moderately weak instruments under a sparse first stage. The divergence between two statistics  depends positively on  parameter $c$. While large values of the first stage coefficients $\Pi_i$ tend to produce large values of both $\frac{\mu^2}{\sqrt{K}}$ and $c$, the relation between the last two is not proportional. A more sparse first stage tends to produce higher values of $c$ (and larger power differences)  for the same level of the identification strength $\frac{\mu^2}{\sqrt{K}}$, and therefore more stark power loss from using the `naive' variance estimator.  Based on a simple simulation design, Figure \ref{fig:power dense} plots the power curves for the leave-one-out AR test with different variance estimators under a sparse first stage (a) and a dense first stage (b). We include additional power comparisons in Section~\ref{section - simulation} and in the Supplementary Appendix.

\section{Pre-test for Weak Identification}\label{section - pretest}

In a prominent paper, Stock and Yogo (2005) introduced a pre-test for weak identification that has gained enormous popularity in applied work. In homoscedastic IV models with small $K$, the concentration parameter fully characterizes the worst bias of the TSLS as a fraction of the OLS bias and the worst rejection rate of TSLS-Wald test.   Stock and Yogo (2005) suggest a set of cut-offs for the first stage $F$ statistic, above which a researcher can guarantee with high (prespecified) probability that the bias of TSLS is not larger than 10\% of the OLS bias, or that the TSLS-Wald statistic does not over-reject by more than 5\%. The cut-offs depend on the goal (bias or size) and the number of instruments. However, these details seem to be mostly disregarded in empirical practice that uses a cut-off of 10, regardless of the goal or the number of instruments.

As with any procedure of such generality, the Stock-Yogo pre-test suffers from multiple drawbacks. First, the pre-test is valid only if the model is homoscedastic. Andrews (2018) shows that  in models calibrated to commonly-used data sets with heteroscedasticity  one may find cases with the first stage $F$ statistics exceeding 1000, that have large over-rejections of the TSLS-Wald test.
Second, the TSLS estimator is less robust to weak identification when $K$ is large. In a homoscedastic model when $K$ is growing proportionally to the sample size, the TSLS estimator is consistent only if $\frac{\pi^\prime Z'Z\pi}{K}\to\infty$, while LIML and \mbox{BTSLS} estimators are consistent when $\frac{\pi^\prime Z'Z\pi}{\sqrt{K}}\to\infty$  as shown in Chao and Swanson (2005).  In this case,  the pre-test becomes too conservative.    Indeed, if  $\frac{\pi^\prime Z'Z\pi}{\sqrt{K}}\to\infty$ but $\frac{\pi^\prime Z'Z\pi}{K}\nrightarrow\infty$, then the pre-test most likely declares weak identification as the expectation of the first stage $F$ equals to $\frac{\pi^\prime Z'Z\pi}{K\sigma_v^2}+1$, even though there exist consistent estimators and a reasonable Wald-test can be constructed.

We propose a new pre-test for weak identification, that allows us to assess the reliability of the JIVE-Wald test.
Our pre-test uses statistic
\begin{equation}\label{eq: F statistics}
\widetilde{F} =\frac{1}{\sqrt{K}\sqrt{\widehat{\Upsilon}}}\sum_{i=1}^{N}\sum_{j\neq i}P_{ij}X_{i}X_{j},
\end{equation}
here $\widehat{\Upsilon}=\frac{2}{K}\sum_{i}\sum_{j\neq i}\frac{P_{ij}^2}{M_{ii}M_{jj}+M_{ij}^2}X_{i}M_{i}XX_{j}M_{j}X$ is an estimate of  the variance  $\Upsilon$ defined in  (\ref{eq: Upsilon}).
The JIVE-Wald test  uses the JIV2 estimator introduced in Angrist et al. (1999):
$$
\widehat{\beta}_{JIVE}=\frac{\sum_{i=1}^{N}\sum_{j\neq i}P_{ij}Y_{i}X_{j}}{\sum_{i=1}^{N}\sum_{j\neq i}P_{ij}X_{i}X_{j}}.
$$
We use the following estimator of the JIVE variance, that is a cross-fit version of the estimator derived in Chao et al. (2012):
$$
\widehat{V}=\frac{\sum_{i=1}^{N}\left(\sum_{j\neq i}P_{ij}X_{j}\right)^2 \frac{\widehat{e}_{i}M_i\widehat{e}}{M_{ii}}+\sum_{i=1}^{N}\sum_{j\neq i}\widetilde{ P}_{ij}^{2}M_iX\widehat{e}_{i}M_jX\widehat{e}_{j}}{\left(\sum_{i=1}^{N}\sum_{j\neq i}P_{ij}X_{i}X_{j}\right)^{2}},
$$
where $\widehat{e}_{i}=Y_{i}-X_{i}\widehat{\beta}_{JIVE}$ and $\widetilde{ P}_{ij}^{2}= \frac{P_{ij}^2}{M_{ii}M_{jj}+M_{ij}^2}.$ The Wald
statistic is defined as
$Wald(\beta_{0})=\frac{\left(\widehat{\beta}_{JIVE}-\beta_{0}\right)^{2}}{\widehat{V}}.$
Our choice of JIVE is based on two considerations. First, according to Hausman et al. (2012), in a heteroscedastic IV model, when $\frac{\pi^\prime Z'Z\pi}{\sqrt{K}}\to\infty$,  LIML and \mbox{BTSLS} become inconsistent, but JIVE is  consistent. Second, the JIVE estimator is a ratio of two quadratic forms similar to the jackknife AR statistic, which motivates the following characterization.

\begin{theorem}\label{thm: first stage F}
Let Assumption \ref{ass: projection matrix} hold for matrix $P$  and  Assumption \ref{ass: errors} hold for  errors $\epsilon_i=(e_i,v_i)'$.
Assume that $\Pi'M\Pi\leq\frac{C \Pi'\Pi}{K}$ and $\frac{\Pi^\prime\Pi}{K^{2/3}} \to 0$ as $N\to\infty$. Then for $\beta=\beta_{0}$,
\begin{align}
\sup_{x,y}\left| \mathbb{P}\left\{ Wald(\beta_{0})\leq x,\widetilde{F}\leq y  \right\}- \mathbb{P}\left\{\frac{\xi^{2}}{1-2\varrho\frac{\xi}{\nu}+\frac{\xi^{2}}{\nu^{2}}}\leq x,\nu\leq y\right\} \right|\to 0
,\label{eq:wald-limit}
\end{align}
where
$\xi$ and $\nu$ are two normal random variables with means $0$ and $\frac{\mu^{2}}{\sqrt{K}\sqrt{\Upsilon}}$, unit variances  and   correlation coefficient $\varrho$  defined in equation (\ref{eq: Upsilon}).
\end{theorem}

Theorem \ref{thm: first stage F} shows that the distribution of the JIVE-Wald statistics can be quite different from its conventional $\chi^2_1$ limit when   $\frac{\mu^{2}}{\sqrt{K}\sqrt{\Upsilon}}$ is small.  If $\frac{\mu^{2}}{\sqrt{K}\sqrt{\Upsilon}}$ is large, then most realizations of the random variable $\nu$ are large as well and the limit of the JIVE-Wald is close to the distribution of $\xi^2$, which is $\chi^2_1$. This suggests that $\frac{\mu^{2}}{\sqrt{K}\sqrt{\Upsilon}}$ is a good measure for identification strength.  The assumption $\frac{\Pi^\prime\Pi}{K^{2/3}} \to 0$ is somewhat restrictive but covers both weakly and strongly identified cases.

Using Theorem \ref{thm: first stage F} we create a pre-test for one definition of weak identification following Stock and Yogo (2005), which stipulates whether the actual size of the conventional 5\% JIVE Wald test could  exceed 10\%. First, we calculate the worst asymptotic rejection rate of  the  JIVE-Wald test for a given theoretical strength of identification $S=\frac{\mu^2}{\sqrt{K}\sqrt{\Upsilon}}$:
\[
R^{\max}_{\alpha}\left(S\right)=\max_{\varrho\in[-1,1]}\mathbb{P}_{S, \varrho}\left\{ \frac{\xi^{2}}{1-2\varrho\frac{\xi}{\nu}+\frac{\xi^{2}}{\nu^{2}}}\geq\chi_{1,1-\alpha}^{2}\right\},
\]
where $\mathbb{P}_{S, \varrho}$ is the probability distribution of $(\xi,\nu)$ as described in Theorem \ref{thm: first stage F}. The quantity $R^{\max}_{\alpha}$ can be straightforwardly obtained from simulations (the maximum rejection occurs at $\varrho=1$). Specifically $S=\frac{\mu^2}{\sqrt{K}\sqrt{\Upsilon}}>2.5 $ implies $R^{\max}_{5\%}\left(S\right)<10\%$.

The strength of identification parameter as measured by $S=\frac{\mu^2}{\sqrt{K}\sqrt{\Upsilon}}$ is unknown in practice.  Theorem~\ref{thm: first stage F} also allows us to construct  a 5\%-test for the null hypothesis that the unknown strength of identification parameter $S=\frac{\mu^2}{\sqrt{K}\sqrt{\Upsilon}} $ is lower than 2.5. This test is based on the statistic $\widetilde{F}$ and rejects whenever $\widetilde{F} > 4.14$. This test is therefore the analog to Stock and Yogo (2005) first stage $F$ pre-test, which tests whether the actual size of the conventional 5\% JIVE Wald test could  exceed 10\%. 

An advantage of  the new pre-test based on $\widetilde{F}$  for weak identification is  that when it is combined with any weak identification robust test, such as our jackknife AR test, to be used when $\widetilde{F}$ is below the cut-off,  we can guarantee that  the size of such two-step procedure is within a tolerance bound of 10\% from the declared nominal size.

\begin{corollary}\label{cor: only one}
Let all assumptions of Theorem \ref{thm: first stage F} hold. Then a two-step test for the null hypothesis $H_0:\beta=\beta_0$ that accepts the null if $\widetilde{F} > 4.14$ and $Wald(\beta_0)<\chi_{1,0.95}^{2}$ or if $\widetilde{F} \leq 4.14$ and $AR(\beta_0)<z_{0.95}$, has an  asymptotic size smaller than 15\%.
\end{corollary}
The attraction of the two-step procedure is that  confidence sets based on the JIVE-Wald test are relatively easy to construct and are well understood by the practitioners.  As we illustrate in simulations, the jackknife AR confidence sets tend  to be wider than the JIVE-Wald confidence sets when identification is strong.  Simulations also suggest the Bonferroni bounds   derived in Corollary \ref{cor: only one} tend to be conservative, as the actual size of the  two-step test does not exceed 7\%.

The 5\% Wald confidence set with 10\% tolerance described in Corollary \ref{cor: only one} is the leading case considered by Stock and Yogo (2005). However, Theorem \ref{thm: first stage F} also allows us to create a two-step procedure with the overall size of  5\% or 10\% by adjusting the cut-off for $\widetilde{F}$  and using Wald and the jackknife AR confidence sets with smaller nominal sizes (and correspondingly larger critical values).  Table~\ref{tab:critical values} tabulates a few  combinations of valid cut-offs and critical values.  As an example of a 5\% two-step procedure, the researcher may compare the $\widetilde{F}$ statistic with 9.98. If $\widetilde{F}$ exceeds the cut-off, the researcher reports a JIVE-Wald confidence set that  uses the 98\% quantile of the $\chi^2_1$ as the critical value. Otherwise, the researcher reports a jackknife AR confidence set that uses the 98\% quantile of the standard normal distribution as the critical value. We apply this procedure to the AK91 example discussed in Section~\ref{section - new} and report the results in Table~\ref{tab:AK91} in italic.

\begin{table}
\begin{centering}
\begin{tabular}{cccc}
\hline
Cut-off for $\widetilde{F}$ & Wald-JIVE   & Jackknife AR   & Overall Size\tabularnewline
\hline
\hline
7.15  & 5.41 (2\%) & 2.32 (1\%) & 5\%\tabularnewline
\hline
9.98  & 5.41 (2\%) & 2.05 (2\%) & 5\%\tabularnewline
\hline
12.86  & 5.41 (2\%) & 1.96 (2.5\%) & 5\%\tabularnewline
\hline
5.01 & 3.84 (5\%) & 2.05 (2\%) & 10\%\tabularnewline
\hline
7.65 & 3.84 (5\%) & 1.75 (4\%) & 10\%\tabularnewline
\hline
\end{tabular}
\par\end{centering}
\caption{\label{tab:critical values}Critical Values for Two-step Procedure}

{\footnotesize{\emph{Notes: }The two-step procedure switches between the Wald-JIVE
test and the jackknife AR test based on the cut-off for $\widetilde{F}$.
When $\widetilde{F}$ is greater than the cut-off, the Wald-JIVE test
is conducted. When $\widetilde{F}$ is less than the cut-off, the
jackknife AR test is conducted. In the parentheses we list the nominal
size associated with the critical values. The last column reports
the overall size for the two-step procedure.}}
\end{table}

\section{Return to Education: Monte Carlo Simulations}\label{section - simulation}

In this section we conduct Monte Carlo simulations to show that the jackknife
AR and the pre-test we develop  are robust to many weak instruments unlike canonical
IV estimators. To maintain the practical relevance, we attempt to preserve the structure of AK91 as described in Section~\ref{section - new}. Specifically, we adopt the simulation design by Angrist and Frandsen (2019).  There is very little endogeneity  in the original AK91, which makes it hard to study the biases of different estimators.  Thus, we follow Angrist and Frandsen (2019) to  introduce additional omitted variable bias to the simulated data.  The simulated data has a nonlinear first stage and is heteroscedastic.   We deviate
from Angrist and Frandsen (2019) in two respects. First, we  vary the
sample size $N$ of the simulated data to be 1.5\%, 1\% and 0.5\%
of the original sample size. This is  to vary  the identification strength. We report the identification strength by  $\frac{\mu^2}{\sqrt{K}\sqrt{\Upsilon}}$ as well as the average $\widetilde{F}$ across simulations. Simulations with sample size equal to 1.5\% of the original sample size produce strong identification in our definition, 1\%  still produce strong identification but close to the weak identification region, while  0.5\%  produce weak identification.When we reduce the sample size we also need to exclude the instruments of the groups that are no longer populated. Second, both in data simulation and in estimation we do not include controls in order to isolate
the implications of many instruments. The Appendix provides more details on our simulation design.

\begin{table}
\begin{centering}
\begin{tabular}{ccccccccccc}
\hline

$N$ & $K$ & Avg. $\widetilde{F}$ & $\frac{\mu^2}{\sqrt{K}\sqrt{\Upsilon}}$ & OLS & 2SLS&2SLS&LIML & LIML &JIVE& JIVE\tabularnewline
&&&& bias & bias&size &bias&size& bias &size \tabularnewline
\hline
\hline
4,923 & 154 & 4.99 & 4.91 & 0.26 & 0.17 &96.6\%& -0.001 &0.6\%& -0.03 &5\%\tabularnewline
\hline
3,209 & 135 & 3.35  & 3.29 & 0.26 & 0.19 &95.7\%& -0.05 &2.7\%& -0.06&5.2\%\tabularnewline
\hline
1,599 & 111 & 1.77 & 1.74 & 0.26 & 0.21 &92.3\%& -0.89 &14.5\%& 1.22& 3.6\%\tabularnewline
\hline
\end{tabular}
\par\end{centering}
\caption{\label{tab:simulation bias} AK91 Simulation
Results: Bias of Different Estimators and Size of Non-robust Tests}
\end{table}

We evaluate the performance of common estimators and tests based on 1000
simulation draws. In Table~\ref{tab:simulation bias}, we report the bias and size of Wald tests based on OLS, 2SLS, LIML and JIVE estimators.  For the Wald test based on the LIML estimator, we calculate the  standard
errors as in Hansen et al. (2008). While Hansen et al. (2008) correct the canonical standard error estimator to be robust to many instruments,
this test is not robust to heteroscedasticity  as LIML itself is inconsistent under heteroscedasticity. For the Wald test based
on the JIVE estimator, we calculate the heteroscedasticity-robust
standard errors as described in Section \ref{section - pretest}.

We find that due to many instruments 2SLS has large bias even under
strong identification. While Hausman et al.
(2012) show LIML is inconsistent under many instruments and heteroscedasticity,
LIML is not too biased in our simulated data, as long as identification is not  weak. We find that JIVE has low bias when identification is strong, but its bias  increases when identification is weak.
The Wald test  based on either LIML or JIVE is not robust to many weak instruments, and we find substantial size distortion for LIML under weak identification.  Surprisingly we do not find large size distortion for JIVE.

In Table \ref{tab:simulation robust size} we
report the rejection frequency of  the robust test we developed in this
paper based on the jackknife AR test statistic. We find that the jackknife AR controls size even under weak identification.
Our proposed pre-test also controls size and is able to switch to
the JIVE-Wald test when identification is strong. In contrast, the first stage F statistics of Stock and Yogo (2005) (FF) are very small even under strong identification, which makes it not very informative.

\begin{table}
\begin{centering}
\begin{tabular}{cccccccc}
\hline
$N$ & $K$ & Avg. FF & Avg. $\widetilde{F}$ & $\frac{\mu^2}{\sqrt{K}\sqrt{\Upsilon}}$ & jackknife AR & pre-test & two-step test\tabularnewline
\hline
\hline
4,923 & 154 & 1.63 & 4.99 & 4.91 & 5.1\% & 70.5\% & 5.8\%\tabularnewline
\hline
3,209 & 135 & 1.44 & 3.35  & 3.29 & 5.6\% & 26.7\% & 6.6\%\tabularnewline
\hline
1,599 & 111 & 1.24 & 1.77 & 1.74 & 6.3\% & 4.5 \% & 7.2\%\tabularnewline
\hline
\end{tabular}
\par\end{centering}
\caption{\label{tab:simulation robust size} AK91 Simulation
Results: Size of Robust Tests}
\end{table}
In Table \ref{tab: length} we compare the length of confidence intervals formed by inverting
 various tests. In particular, when identification is strong, jackknife AR confidence sets are longer (less efficient) but
are not unreasonably long compared to the Wald tests based on  LIML and JIVE. In this case, a pre-test can improve the efficiency by switching to the Wald test based on JIVE.
As with the canonical AR
test, the jackknife AR test can result in confidence intervals with
infinite length. We report the probability of infinite length in the last column of   Table \ref{tab: length}, and note that  such probability increases as identification
gets weaker.
\begin{table}
\begin{centering}
\begin{tabular}{ccccccccc}
\hline
$N$ & $K$ & Avg. $\widetilde{F}$ & $\frac{\mu^2}{\sqrt{K}\sqrt{\Upsilon}}$ & 2SLS & LIML & JIVE & jackknife AR & infinite jackknife AR\tabularnewline
\hline
\hline
4,923 & 154 & 4.99 & 4.91 & 0.18 & 1.14& 0.81 & 1.66 & 1\% \tabularnewline
\hline
3,209 & 135 & 3.35 & 3.29 & 0.20 & 1.23 & 1.41 & 2.77 & 11\%\tabularnewline
\hline
1,599 & 111 & 1.77  & 1.74 & 0.24 & 1.46 & 5244 & 6.90 & 49.6\% \tabularnewline
\hline
\end{tabular}
\par\end{centering}
\caption{\label{tab: length} AK91 Simulation Results, Length of Confidence
Interval}
\end{table}

\begin{figure}
     \centering
     \begin{subfigure}[t]{0.45\textwidth}
         \centering
         \includegraphics[width=\textwidth]{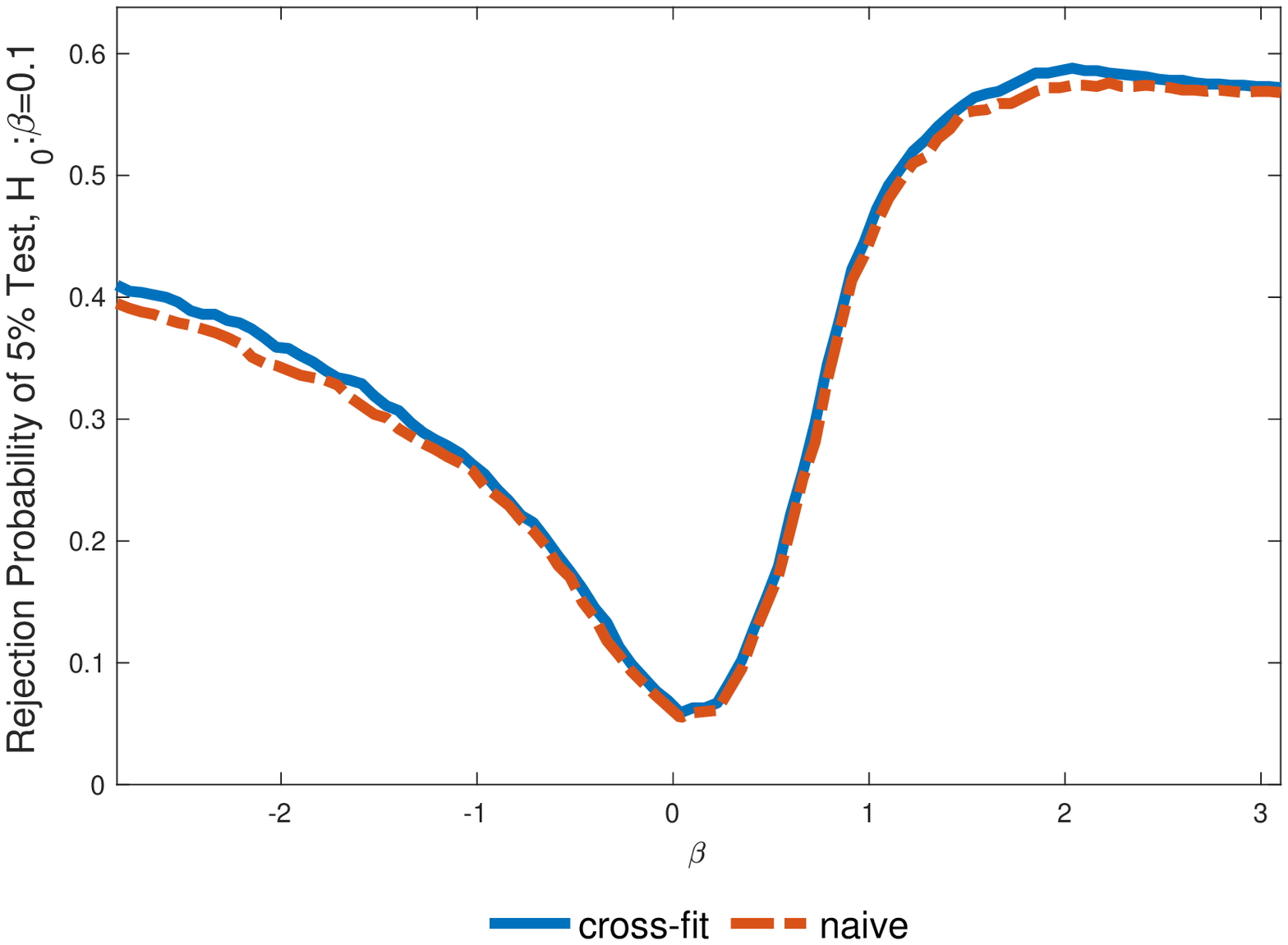}
         \caption{$N=1599,K=111$}
     \end{subfigure}
     ~
     \begin{subfigure}[t]{0.45\textwidth}
         \centering
         \includegraphics[width=\textwidth]{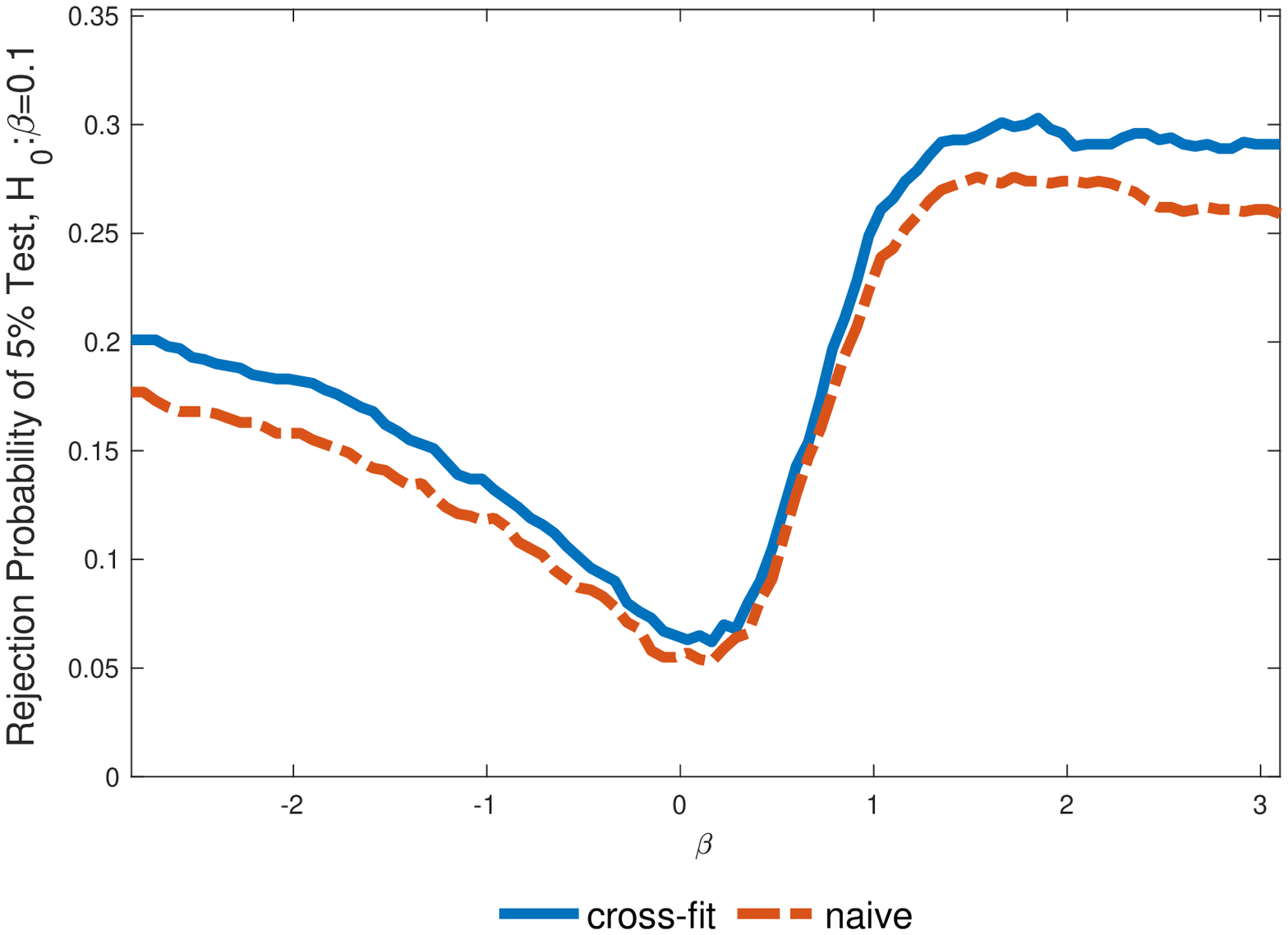}
         \caption{$N=796,K=77$}
     \end{subfigure}

\caption{Power curves with varying identification strength, $H_{0}:\beta=0.1$}
        \label{fig:power}
\end{figure}

To complement the discussion in Section~\ref{subsection - power}, we compare the performance of the jackknife AR test based on our proposed ``cross-fit'' variance estimator with that based on the ``naive'' variance estimator. Since power loss does not show up with strong identification, we further reduce the sample size to be 0.25\% of the original size. In Table~\ref{tab:var comp} we confirm that the size is not affected by the choice of variance estimator. Figure~\ref{fig:power} demonstrates the difference in power for the jackknife AR tests with the cross-fit and the naive variance estimators. The ``cross-fit'' variance estimator performs slightly better in terms of power when identification is weak. As shown in the last two columns of Table \ref{tab:var comp}, the power difference is also reflected in fewer unbounded confidence intervals based on the jackknife AR test, and shorter confidence intervals when the bounded using the ``cross-fit" variance estimator.

\begin{table}
\begin{subtable}[t]{\linewidth}
\caption{``cross-fit'' variance estimator}
\begin{tabular}{cccccccc}
\hline
$N$ & $K$ & Avg. $\widetilde{F}$ & $\frac{\mu^{2}}{\sqrt{K}\sqrt{\Upsilon}}$ & jackknife AR size & two-step test & CI length & infinite CI\tabularnewline
\hline
\hline
4,923 & 154 & 4.99 & 4.91 & 5.1\% & 5.8\% & 1.66 & 1\%\tabularnewline
\hline
3,209 & 135 & 3.35 & 3.29 & 5.6\% & 6.6\% & 2.77 & 11\%\tabularnewline
\hline
1,599 & 111 & 1.77 & 1.74 & 6.3\% & 7.2\% & 6.90 & 49.6\%\tabularnewline
\hline
796 & 77 &  0.92 & 0.91 & 6.5\%  & 6.5\% & 10.26 & 74.4\%\tabularnewline
\hline
\end{tabular}
\end{subtable}

\begin{subtable}[t]{\linewidth}
\caption{``naive'' variance estimator}
\begin{tabular}{cccccccc}
\hline
$N$ & $K$ & Avg. $\widetilde{F}$ & $\frac{\mu^{2}}{\sqrt{K}\sqrt{\Upsilon}}$ & jackknife AR size & two-step test & CI length & infinite CI\tabularnewline
\hline
\hline
4,923 & 154 & 4.99 & 4.91 & 4.9\% & 5.7\% & 1.81 & 1\%\tabularnewline
\hline
3,209 & 135 & 3.35 & 3.29 & 5.4\% & 6.5\% & 2.99 & 11.1\%\tabularnewline
\hline
1,599 & 111 & 1.77 & 1.74 & 5.9\% & 6.8\% & 6.95 & 51.1\%\tabularnewline
\hline
796 & 77 &  0.92 & 0.91 & 5.4\%  & 5.4\% & 8.86 & 77.3\%\tabularnewline
\hline
\end{tabular}
\end{subtable}
\caption{\label{tab:var comp}Angrist and Krueger (1991) Simulation Results,
Comparison of Variance Estimation}
\end{table}

\section{Conclusion}\label{section- conclusion}
In this paper, we focus on identification for linear IV models with many instruments.  In this environment, we characterize weak identification as a situation where an analog of the concentration parameter stays bounded relative to the square root of the number of instruments in large samples. We introduce a jackknifed version of the AR test that is robust to  our definition of weak identification and heteroscedasticity. We also propose a pre-test for weak identification and correspondingly a two-step testing procedure in the spirit of Stock and Yogo (2005).  Unlike the pre-test proposed by Stock and Yogo (2005), our two-step test controls size distortion even under heteroscedasticity and with  many instruments.  As an empirical example, our pre-test rejects weak identification in  Angrist and Krueger (1991) where up to 1,530 instruments are used.

\subsection*{Data Availability Statement}
The data underlying this article are available in ``Replication package for Inference with Many Weak Instruments", at \url{https://doi.org/10.5281/zenodo.5546157}.

\section*{References}

Anatolyev, S., and Gospodinov, N. (2011). ``Specification Testing in
Models with Many Instruments." \emph{Econometric Theory} 27, 427\textendash 441. \\
Andrews, I. (2018). ``Valid Two-Step Identification-Robust Confidence Sets for GMM." \emph{The Review of Economics and Statistics}, 100, 337\textendash348. \emph{Supplementary Appendix} \\
Andrews, D.W.K., and Stock, J.H. (2007). ``Testing with many weak instruments.''
\emph{Journal of Econometrics} 138, 24\textendash 46. \\
Andrews D, Moreira M, Stock J. (2006). ``Optimal two-sided invariant similar
tests of instrumental variables regression." \emph{Econometrica} 74:715\textendash 752 \\
Angrist, J.D., and Frandsen B. (2019). ``Machine Labor" \emph{NBER working paper} 26584.\\
Angrist, J.D., Imbens, G.W., and Krueger, A.B. (1999). ``Jackknife instrumental
variables estimation.'' \emph{Journal of Applied Econometrics} 14, 57\textendash67. \\
Angrist, J.D., and Krueger, A.B. (1991). ``Does Compulsory School Attendance Affect Schooling and Earnings?"  \emph{The Quarterly Journal of Economics} 106, 979\textendash1014.\\
Bekker, P.A. (1994). ``Alternative Approximations to the Distributions of Instrumental Variable Estimators.'' \emph{Econometrica} 62, 657\textendash681.\\
Bhuller, M., Dahl, G.B., Loken, K.V. and M. Mogstad (2020): ``Incarceration, Recidivism, and Employment,'' \emph{Journal of Political Economy}, 128(4), 1269\textendash1324.\\
Bound, J., Jaeger, D.A. and  Baker, R.M. (1995) ``Problems with Instrumental Variables Estimation when the Correlation between the Instruments and the Endogenous Explanatory Variable is Weak,'' \emph{Journal of the American Statistical Association}, 90:430, 443\textendash450.\\
Chao, J.C., Hausman, J.A., Newey, W.K., Swanson, N.R., and Woutersen, T. (2014). ``Testing overidentifying restrictions with many instruments and heteroskedasticity.'' \emph{Journal of Econometrics} 178, 15\textendash21.\\
Chao, J.C., and Swanson, N.R. (2005). ``Consistent Estimation with a Large Number of Weak Instruments." \emph{Econometrica} 73, 1673\textendash1692.\\
Chao, J.C., Swanson, N.R., Hausman, J.A., Newey, W.K., and Woutersen,
T. (2012). ``Asymptotic Distribution of JIV in a heteroscedastic IV
Regression with Many Instruments.'' \emph{Econometric Theory} 28, 42\textendash 86. \\
Crudu, F., Mellace, G., and Sandor, Z. (2020). ``Inference in Instrumental Variables Models with Heteroskedasticity and Many Instruments.'' \emph{Econometric Theory}, forthcoming.\\
Dobbie, W., Goldin, J., and Yang, C.S. (2018). ``The Effects of Pretrial Detention on Conviction, Future Crime, and Employment: Evidence from Randomly Assigned Judges." \emph{American Economic Review} 108, 201\textendash240.\\
Fama, E.F., and MacBeth, J.D. (1973). ``Risk, Return, and Equilibrium: Empirical Tests." \emph{Journal of Political Economy} 81, 607\textendash636.\\
Hansen, C., Hausman, J., and Newey, W. (2008). ``Estimation With Many Instrumental Variables." \emph{Journal of Business \& Economic Statistics} 26, 398\textendash 422.\\
Hausman, J.A., Newey, W.K., Woutersen, T., Chao, J.C., and Swanson,
N.R. (2012). ``Instrumental variable estimation with heteroscedasticity
and many instruments.'' \emph{Quantitative Economics} 3, 211\textendash 255. \\
Kleibergen F. (2002). ``Pivotal statistics for testing structural parameters
in instrumental variables regression.'' \emph{Econometrica} 70:1781\textendash 1803. \\
Kline, P., Saggio, R., and S{\o}lvsten, M. (2020). ``Leave-out estimation
of variance components.'' \emph{Econometrica} 88, 1859\textendash1898.\\
Maestas, N., Mullen, K.J., and Strand, A. (2013). ``Does Disability Insurance Receipt Discourage Work? Using Examiner Assignment to Estimate Causal Effects of SSDI Receipt.'' \emph{American Economic Review} 103, 1797\textendash1829.\\
M{\"u}ller, U. K. (2011). ``Efficient Tests Under a Weak Convergence Assumption.'' \emph{Econometrica} 79 (2): 395\textendash435.  \\
Newey, W. (2004). ``Many Instrument Asymptotics.'' \\
Newey, W.K., and Robins, J.R. (2018). ``Cross-Fitting and Fast Remainder
Rates for Semiparametric Estimation.'' \\
Newey, W.K., and Windmeijer, F. (2009). ``Generalized Method of Moments
With Many Weak Moment Conditions.'' \emph{Econometrica} 77, 687\textendash 719. \\
Sampat, B., and Williams, H.L. (2019). ``How Do Patents Affect Follow-On Innovation? Evidence from the Human Genome.''  \emph{American Economic Review} 109, 203\textendash236.\\
Shanken, J. (1992). ``On the Estimation of Beta-Pricing Models." \emph{The Review of Financial Studies} 5, 1\textendash33.\\
Staiger, D., and Stock, J.H. (1997). ``Instrumental Variables Regression with Weak Instruments.'' \emph{Econometrica} 65 (3): 557\textendash86. \\
Stock, J.H., and Yogo, M. (2005). ``Testing for weak instruments in Linear IV regression. In Identification and Inference for Econometric Models: Essays in Honor of Thomas Rothenberg," pp. 80\textendash108.\\
%\pagebreak
\section{Appendix with Proofs}
Let $C$  be a universal constant (that may be different in different lines but does not depend on $N$ or $K$).

\paragraph{Proof of Theorem \ref{thm: negative result}.} Denote
 $                      A $ to be an upper-triangular matrix, such that
 $A\Omega A'=I_2$. The sufficient statistic in model (\ref{eq: iv model}) is
\begin{equation}\label{eq: sufficient statistics}
\left(
  \begin{array}{c}
    \xi_1 \\
    \xi_2 \\
  \end{array}
\right)=(A\otimes I_K)\cdot
\left(
  \begin{array}{c}
    (Z'Z)^{-1/2}Z'Y \\
    (Z'Z)^{-1/2}Z'X \\
  \end{array}
\right)\sim N\left(\left(
                     \begin{array}{c}
                       \widetilde{\beta}\Pi\\
                       \Pi\\
                     \end{array}
                   \right),      I_{2K}
\right)
\end{equation}
where $\widetilde{\beta}=(1,0)A(\beta,1)'$ is a (known) linear one-to-one transformation of $\beta$. Denote the corresponding null and alternative as $\widetilde\beta_0$ and $\widetilde\beta^*$. We denote also $\Pi=\frac{(Z'Z)^{1/2}\pi }{\sigma_v}$, which is one-to-one transformation of $\pi$.
It is enough to restrict attention to the tests that depend on the data through sufficient statistics only. Indeed, for any test $\psi\in\Psi_N$ we may construct a test $\psi^S=\mathbb{E}(\psi|\xi_1,\xi_2)$ which depends on the data only through the sufficient statistics. Due to the law of iterated expectations the size and the power of $\psi^S$ is the same as the initial $\psi$.

Let $\mathcal{U}$ be the group of rotations on $\mathbb{R}^K$, that is $U\in \mathcal{U}$ are such $U'U=I_K$.  Notice that the model is invariant to group $\mathcal{U}$, namely if $(\xi_1,\xi_2)$ satisfy model (\ref{eq: sufficient statistics}) with parameters $(\widetilde\beta, \Pi)$ then $(U\xi_1,U\xi_2)$ satisfy model (\ref{eq: sufficient statistics}) with parameters $(\widetilde\beta, U\Pi)$. Note that $\Pi'\Pi=(U\Pi)'(U\Pi)$. This implies that for any function $f$ we have $\mathbb{E}_{(\widetilde\beta,\Pi)}f(U\xi_1,U\xi_2)=\mathbb{E}_{(\widetilde\beta,U\Pi)}f(\xi_1,\xi_2)$.

We call a test $\psi=\psi(\xi_1,\xi_2)$ invariant to rotations iff for any $U\in \mathcal{U}$ we have $\psi(U\xi_1,U\xi_2)=\psi(\xi_1,\xi_2)$ for all realizations of $(\xi_1,\xi_2)$. The maximum in Theorem \ref{thm: negative result} is achieved at an invariant test. Indeed, take any test $\psi\in \Psi_N$ that has size $\alpha$, that is, $\mathbb{E}_{(\widetilde\beta_0,\Pi)}\psi(\xi_1,\xi_2)\leq \alpha$ for all $\Pi$.
Let us consider a new test
$
\psi^*(\xi_1,\xi_2)=\int_{U\in\mathcal{U}}\psi(U\xi_1,U\xi_2)dU,
$
where the integral is taken uniformly over the unit sphere in $\mathbb{R}^{K}$.
By construction, $\psi^*$ is an invariant test as for any $\widetilde{U}\in \mathcal{U}$, we have $U\widetilde{U}\in\mathcal{U}$ for all $U\in\mathcal{U}$ so that
\begin{align*}
\psi^*(\widetilde{U}\xi_1,\widetilde{U}\xi_2)&=\int_{U\in\mathcal{U}} \psi(U\widetilde{U}\xi_1,U\widetilde{U}\xi_2)dU=\int_{U\in\mathcal{U}}\psi(U\xi_1,U\xi_2)dU.
\\
\mathbb{E}_{(\widetilde\beta_0,\Pi)}\psi^*(\xi_1,\xi_2) &= \int_{U\in\mathcal{U}}\left\{\mathbb{E}_{(\widetilde\beta_0,\Pi)}\psi(U\xi_1,U\xi_2)\right\}dU
=\int_{U\in\mathcal{U}}\left\{\mathbb{E}_{(\widetilde\beta_0,U\Pi)}\psi(\xi_1,\xi_2)\right\}dU\leq \alpha.
\end{align*}
So, it has correct size. Now we check that the minimal power of $\psi^*$ achieved over alternatives $(\widetilde\beta^*,\Pi)$ with $\Pi$ such that $\frac{\Pi'\Pi}{\sqrt{K}}=C$ is not smaller than that of $\psi$.
Assume that the minimum of power for test $\psi$ is achieved at the alternative $\Pi^*$: $\min_{\frac{\Pi'\Pi}{\sqrt{K}}=C}\mathbb{E}_{(\widetilde\beta^*,\Pi)}\psi(\xi_1,\xi_2)= \mathbb{E}_{(\widetilde\beta^*,\Pi^*)}\psi(\xi_1,\xi_2).$
Then, similarly to above:
\begin{align*}
\min_{\frac{\Pi'\Pi}{\sqrt{K}}=C}\mathbb{E}_{(\widetilde\beta^*,\Pi)}\psi^*(\xi_1,\xi_2)= \min_{\frac{\Pi'\Pi}{\sqrt{K}}=C}\int_{U\in\mathcal{U}} \left\{\mathbb{E}_{(\widetilde\beta^*,U\Pi)}\psi(\xi_1,\xi_2)\right\}dU\geq\\ \geq\int_{U\in\mathcal{U}}\min_{\frac{\Pi'\Pi}{\sqrt{K}}=C} \left\{\mathbb{E}_{(\widetilde\beta^*,U\Pi)}\psi(\xi_1,\xi_2)\right\}dU=\mathbb{E}_{(\widetilde\beta^*,\Pi^*)}\psi(\xi_1,\xi_2).
\end{align*}
All invariant tests depend on the data only through maximal invariant. Thus, we should only consider tests that depend on the data through  statistics $Q=(Q_1,Q_2,Q_3)=(\xi_1'\xi_1, \xi_1'\xi_2, \xi_2'\xi_2)$.
If $\Pi'\Pi/\sqrt{K}\to C$ then $Q$ converges to the following distribution:
\begin{equation}\label{eq: limit experiment}
\left(
  \begin{array}{c}
    \frac{\xi_1'\xi_1-K}{\sqrt{2K}} \\
    \frac{\xi_1'\xi_2}{\sqrt{K}} \\
    \frac{\xi_2'\xi_2-K}{\sqrt{2K}} \\
  \end{array}
\right)\Rightarrow N\left(
\left(
  \begin{array}{c}
    \widetilde\beta^2\frac{C}{\sqrt{2}} \\
    \widetilde\beta C \\
    \frac{C}{\sqrt{2}} \\
  \end{array}
\right)
,I_3\right)=\left(
              \begin{array}{c}
                Q_{\infty, 1} \\
                Q_{\infty, 2} \\
                Q_{\infty, 3} \\
              \end{array}
            \right) = Q_{\infty}.
\end{equation}
According to Theorem 1 of M{\"u}eller (2011) the limit of the maximal power of tests in experiment based on $Q$ is bounded above  by the maximal power achieved in the limit experiment described on $ Q_{\infty}$ as defined in the right hand side of equation (\ref{eq: limit experiment}).
Notice that the maximal achievable power $\mathbb{E}_{\widetilde\beta^*,C}\psi^*(Q_\infty)$ is strictly less than 1 for any fixed $\beta^*$ and fixed $C$. Indeed, the best achievable power in the limit experiment
(\ref{eq: limit experiment})
is no more than the best achievable power in the  experiment when $C$ is known. If $C$ is known, the optimal test follows from the Neyman-Pearson lemma, and its power is less than 1.

\paragraph{Proof of Theorem \ref{thm: consistency of variance null}.}Assumptions \ref{ass: projection matrix} and \ref{ass: errors} imply
$$
1\geq\frac{1}{K}\sum_{i}\sum_{j\neq i}P_{ij}^2=\frac{1}{K}\sum_{i}\sum_{j}P_{ij}^2-\frac{1}{K}\sum_iP_{ii}^2\geq 1-\delta\frac{1}{K}\sum_iP_{ii}=1-\delta.
$$
Thus, $(1-\delta)(c^*)^2<\Phi<(C^*)^2$ and it is sufficient to prove that $\widehat\Phi-\Phi\to^p0$. The last statement holds due to Lemma \ref{lem: consistency of variance null} applied to $\xi_i=(e_i,e_i,e_i)'.$ $\Box$

\begin{lemma}\label{lem: consistency of variance null}
Let Assumption \ref{ass: projection matrix} hold.  Assume the errors $\xi_i=(\xi_{i}^{(1)},\xi_{i}^{(2)},\xi_{i}^{(3)})'$ are independent mean zero random vectors with $\max_i\mathbb{E}\|\xi_i\|^6<C$. Then as    $ N\to\infty,$ we have:
\begin{align*}
\frac{1}{K}\sum_{i}\sum_{j\neq i}\left\{\frac{P_{ij}^2}{M_{ii}M_{jj}+M_{ij}^2} \left[\xi_{i}^{(1)}M_i\xi^{(2)}\right]\left[\xi_{j}^{(1)}M_j\xi^{(3)}\right]-P_{ij}^2 \mathbb{E}\left[\xi_{i}^{(1)}\xi_i^{(2)}\right]\mathbb{E}\left[\xi_{j}^{(1)}\xi_j^{(3)}\right]\right\}\to^p 0.
\end{align*}
\end{lemma}

\paragraph{Proof of Lemma \ref{lem: consistency of variance null}.} Here we use the following notation $\widetilde{P}_{ij}^2=\frac{P_{ij}^2}{M_{ii}M_{jj}+M_{ij}^2}$. Notice that
\begin{align*}
\frac{1}{K}\sum_{i}\sum_{j\neq i} \widetilde{P}_{ij}^2 \mathbb{E} \left[\xi_{i}^{(1)}M_i\xi^{(2)}\xi_{j}^{(1)}M_j\xi^{(3)}\right] =\frac{1}{K} \sum_{i}\sum_{j\neq i}\widetilde{P}_{ij}^2 M_{ii}M_{jj} \mathbb{E}\left[\xi_{i}^{(1)}\xi_i^{(2)}\xi_{j}^{(1)}\xi_j^{(3)}\right]+\\+ \frac{1}{K}\sum_{i}\sum_{j\neq i}\widetilde{P}_{ij}^2 M_{ij}^2 \mathbb{E}\left[\xi_{i}^{(1)}\xi_j^{(2)}\xi_{j}^{(1)}\xi_i^{(3)}\right]=
\frac{1}{K}\sum_{i}\sum_{j\neq i}P_{ij}^2 \mathbb{E}\left[\xi_{i}^{(1)}\xi_i^{(2)}\right]\mathbb{E}\left[\xi_{j}^{(1)}\xi_j^{(3)}\right]
\end{align*}
Define
$
\xi_{ij}=\xi_{i}^{(1)}M_i\xi^{(2)}\xi_{j}^{(1)}M_j\xi^{(3)}- \mathbb{E}\left[\xi_{i}^{(1)}M_i\xi^{(2)}\xi_{j}^{(1)}M_j\xi^{(3)}\right],
$
then we need to prove that
$
\frac{1}{K}\sum_{i}\sum_{j\neq i}\widetilde{P}_{ij}^2\xi_{ij}\to^p 0
$.
Since $\frac{1}{K}\sum_{i}\sum_{j\neq i}\widetilde{P}_{ij}^2\xi_{ij}$ has zero mean, it  is sufficient to show that the variance of each term in expression (\ref{eq 2}) defined below converges to zero (here $I_4$ is a summation over distinct indexes $(i,i',j,j')$):
\begin{align}\notag
&\mathbb{E}\left(\frac{1}{K}\sum_{i}\sum_{j\neq i}\widetilde{P}_{ij}^2\xi_{ij}\right)^2=\frac{1}{K^2}\sum_{i}\sum_{j\neq i}\widetilde{P}_{ij}^4\mathbb{E}\xi_{ij}^2+\\+&\frac{1}{K^2}\sum_{i}\sum_{j\neq i}\sum_{i^\prime\neq \{i,j\}}
\widetilde{P}_{ij}^2\widetilde{P}_{ii^\prime}^2\mathbb{E}\xi_{ij}\xi_{ii^\prime}+\frac{1}{K^2} \sum_{I_4}\widetilde{P}_{ij}^2\widetilde{P}_{i'j'}^2\mathbb{E}\xi_{ij}\xi_{i'j'}.\label{eq 2}
\end{align}
First, we prove that $\max_{i,j}\mathbb{E}\xi_{ij}^2< C$. We expand $\xi_{ij}=
A_{1,ij}+A_{2,ij}+A_{3,ij}$, where:
\begin{align*}
A_{1,ij}=& M_{ii}M_{jj}\left(\xi_{i}^{(1)}\xi_{i}^{(2)}\xi_{j}^{(1)}\xi_{j}^{(3)} -\mathbb{E}[\xi_{i}^{(1)}\xi_{i}^{(2)}\xi_{j}^{(1)}\xi_{j}^{(3)}]\right)+ M_{ij}^{2}\left(\xi_{i}^{(1)}\xi_{i}^{(3)}\xi_{j}^{(1)}\xi_{j}^{(2)}- \mathbb{E}[\xi_{i}^{(1)}\xi_{i}^{(3)}\xi_{j}^{(1)}\xi_{j}^{(2)}]\right),\\
A_{2,ij}=&\xi_{i}^{(1)}\xi_{j}^{(1)}\sum_{i'\neq\{i,j\}}\left(M_{ii}M_{ji'}\xi_{i}^{(2)}\xi_{i'}^{(3)}+ M_{ii'}M_{ij}\xi_{i'}^{(2)}\xi_{i}^{(3)}+M_{jj}M_{ii'}\xi_{i'}^{(2)}\xi_{j}^{(3)}+ M_{ji'}M_{ij}\xi_{j}^{(2)}\xi_{i'}^{(3)}\right),\\
A_{3,ij}=&\xi_{i}^{(1)}\xi_{j}^{(1)} \sum_{i'\neq\{i,j\}}\sum_{j'\neq\{i,j\}}M_{ii'}M_{jj'}\xi^{(2)}_{i'}\xi^{(3)}_{j'}.
\end{align*}
It is sufficient to show that $\max_{i,j}\mathbb{E}A^2_{s,ij}$ is bounded for all $s=1,2,3$. The moment condition implies
$
\mathbb{E}A_{1,ij}^2\leq  C\left(M_{ii}M_{jj}+M_{ij}^2\right)^2 \leq C.$
Below we use  that non-zero correlations between summands in $A_{s,ij}$ imply that some indexes  must coincide. We also use  Lemma S1.1 from the Supplementary Appendix:
\begin{align*}
\mathbb{E}A_{2,ij}^2\leq C\sum_{i'}(M_{ii}M_{ji'}+ M_{ii'}M_{ij}+M_{jj}M_{ii'}+M_{ji'}M_{ij})^2\leq C,
\\
\mathbb{E}A_{3,ij}^2\leq  C\sum_{i'\neq\{i,j\}}\sum_{j'\neq\{i,j\}}\left( P_{ii'}^2P_{jj'}^2+|P_{ii'}P_{jj'}P_{ij'}P_{ji'}|\right)\leq C.
\end{align*}
Next notice that
\begin{equation}\label{eq: bound on Pij}
  \widetilde{P}_{ij}^2=\frac{P_{ij}^2}{M_{ii}M_{jj}+M_{ij}^2}\leq \frac{P_{ij}^2}{(1-P_{ii})(1-P_{jj})}\leq \frac{1}{(1-\delta)^2}P_{ij}^2.
\end{equation}
Lemma B1 in Chao et al (2012)
gives that $\sum_{i}\sum_{j\neq i}P_{ij}^4\leq K$ and $\sum_{i}\sum_{j\neq i}\sum_{j^\prime\neq i,j^\prime\neq j}
P_{ij}^2P_{ij^\prime}^2\leq K$.
Thus, given the bound on $\max_{i,j}E\xi_{ij}^2< C$ and by Cauchy-Schwarz inequality $\max_{i,j,k}|\mathbb{E}\xi_{ij}\xi_{ik}|<C$, the first two terms in expression (\ref{eq 2})  converge to zero.

For the last term in (\ref{eq 2}), since $i,i',j,j'$ are all distinct,  we have $\mathbb{E}A_{1,ij}A_{s,i'j'}=0$ for $s=2,3$, and $\mathbb{E}A_{2,ij}A_{3,i'j'}=0$. The non-zero terms in $\mathbb{E}\xi_{ij}\xi_{i'j'}$ are
\begin{align*}
\left|\mathbb{E}A_{2,ij}A_{2,i'j'}\right|\leq &C\left|(M_{ii}M_{jj'}+M_{ij}M_{ij'})
(M_{i'i'}M_{jj'}+M_{i'j}M_{i'j'})\right|+\\+&C\left|(M_{jj}M_{ii'}+M_{ji'}M_{ij})(M_{j'j'}M_{ii'}+ M_{j'i'}M_{ij'})\right|.
\\
\left| \mathbb{E}A_{3,ij}A_{3,i'j'}\right|\leq &C (P_{ii'}P_{jj'}+P_{ij'}P_{i'j})^2.
\end{align*}
Given  inequality (\ref{eq: bound on Pij}) and the symmetry of summation, and statements (a)-(e) proved in Lemma S1.2 in the Supplementary Appendix, we obtain that the last two terms in equation (\ref{eq 2}) converge to zero.  $\Box$

\paragraph{Proof of Theorem \ref{thm: consistency of variance alternative}.}
Denote $\lambda_i=M_i\Pi$, then
\begin{align*}
\widehat\Phi=\frac{2}{K}\sum_{i}\sum_{j\neq i}\widetilde{P}_{ij}^2\left(\eta_i+\Delta\Pi_i\right)\left(M_i\eta+\Delta\lambda_i\right)
\left(\eta_j+\Delta\Pi_j\right)\left(M_j\eta+\Delta\lambda_j\right).
\end{align*}
Let us define $\widehat{\Phi}_{0}=\frac{1}{K}\sum_{i}\sum_{j\neq i}\widetilde{P}_{ij}^{2}\eta_{i}M_{i}\eta \eta_{j}M_{j}\eta$. Assumption \ref{ass: errors}  guarantees that the variance of $\eta_i=e_i+\Delta\cdot v_i$ is uniformly bounded. Lemma \ref{lem: consistency of variance null} with $\xi_i=(\eta_i,\eta_i,\eta_i)'$ gives
$
\left|\widehat{\Phi}_{0}-\Phi\right|\to^p 0
$ uniformly over bounded $\Delta$. Lemma \ref{lem: on estimation under alternative} with $\xi_i=(\eta_i,\eta_i,\eta_i,\eta_i)'$ implies $\widehat\Phi-\widehat{\Phi}_{0}\to^p0.$
$\Box$
\begin{lemma}\label{lem: on estimation under alternative}
Let $\xi_i=(\xi_{i}^{(1)},\xi_{i}^{(2)},\xi_{i}^{(3)},\xi_{i}^{(4)})'$ be independent mean zero  $4\times 1$ random vectors, such that $\mathbb{E}\|\xi_i\|^4< C.$ Let Assumption \ref{ass: projection matrix} hold. Assume that $\lambda'\lambda\leq\frac{C}{K}\Pi'\Pi$ and  $\Delta^2\cdot\frac{\Pi'\Pi}{K}\to 0$ as $N\to\infty$. Then
\begin{align*}
\frac{1}{K}\sum_{i}\sum_{j\neq i}\widetilde{P}_{ij}^2\left(\xi_{i}^{(1)}+\Delta\Pi_i\right)\left(M_i\xi^{(2)}+\Delta\lambda_i\right)
\left(\xi_{j}^{(3)}+\Delta\Pi_j\right)\left(M_j\xi^{(4)}+\Delta\lambda_j\right)-\\- \frac{1}{K}\sum_{i}\sum_{j\neq i}\widetilde{P}_{ij}^2\xi_{i}^{(1)}M_i\xi^{(2)}
\xi_{j}^{(3)}M_j\xi^{(4)}\to^p0.
\end{align*}
\end{lemma}
\paragraph{Proof of Lemma \ref{lem: on estimation under alternative}.}
We write the main expression of interest as a polynomial of fourth power in $\Delta$: $\Delta^4A_4+\Delta^3A_3+\Delta^2A_2+\Delta A_1$ and prove that all terms are negligible $\Delta^lA_l\to^p0$ by showing that their means and variances converge to zero.  Notice that for expressions with  identical structure but different components of $\xi_i$, the proof of their negligibility is exactly the same. Thus for simplicity we abuse the notation and drop the superscripts to $\xi_i$ when we can consolidate these expressions.  For example, we write the expression for one of the terms in $A_3$  as  $\frac{1}{K}\sum_{i}\sum_{j\neq i}\widetilde{P}_{ij}^2\Pi_i\lambda_i\lambda_j\xi_j$, which collects both $\frac{1}{K}\sum_{i}\sum_{j\neq i}\widetilde{P}_{ij}^2\Pi_i\lambda_i\lambda_j\xi_j^{(1)}$ and $\frac{1}{K}\sum_{i}\sum_{j\neq i}\widetilde{P}_{ij}^2\Pi_i\lambda_i\lambda_j\xi_j^{(3)}$.  We also treat $\xi_i$ in all expressions below as scalar.
\begin{align*}
A_4&=\frac{1}{K}\sum_{i}\sum_{j\neq i}\widetilde{P}_{ij}^2\Pi_i\lambda_i\Pi_j\lambda_j;\\
A_3&=\frac{1}{K}\sum_{i}\sum_{j\neq i}\widetilde{P}_{ij}^2\Pi_i\lambda_i\lambda_j\xi_j+\frac{1}{K}\sum_{i}\sum_{j\neq i}\widetilde{P}_{ij}^2\Pi_i\lambda_i\Pi_jM_j\xi;
\\
A_2&=\frac{1}{K}\sum_{i}\sum_{j\neq i}\widetilde{P}_{ij}^2\lambda_i\lambda_j\xi_i\xi_j+\frac{1}{K}\sum_{i}\sum_{j\neq i}\widetilde{P}_{ij}^2\lambda_i\xi_i\Pi_jM_j\xi+\\
&+\frac{1}{K}\sum_{i}\sum_{j\neq i}\widetilde{P}_{ij}^2\lambda_i\Pi_i\xi_j M_j\xi+\frac{1}{K}\sum_{i}\sum_{j\neq i}\widetilde{P}_{ij}^2\Pi_i\Pi_jM_i\xi M_j\xi;\\
A_1&=\frac{1}{K}\sum_{i}\sum_{j\neq i}\widetilde{P}_{ij}^2\lambda_i\xi_iM_j\xi\xi_j+\frac{1}{K}\sum_{i}\sum_{j\neq i}\widetilde{P}_{ij}^2\Pi_iM_i\xi \xi_jM_j\xi.
\end{align*}
Term $A_4$ is deterministic.  We use  bound (\ref{eq: bound on Pij}) and Lemma S1.3 (d):
$$
\Delta^4|A_4|\leq \frac{C\Delta^4\Pi'\Pi\lambda'\lambda}{K}\leq \frac{C\Delta^4(\Pi'\Pi)^2}{K^2}\to 0.
$$
Term $A_3$ is mean zero. Using the inequality $Var(X+Y)\leq 2 Var(X)+2Var(Y)$ we have:
\begin{align*}
\Delta^6Var(A_3)\leq\frac{C\Delta^6}{K^2}\left(\sum_j\left(\sum_iP_{ij}^2|\Pi_i||\lambda_i|\right)^2\lambda_j^2 +\sum_k\left(\sum_{i}\sum_{j\neq i}\widetilde{P}_{ij}^2\Pi_i\lambda_i\Pi_jM_{jk}\right)^2\right)\leq\\ \leq\frac{C\Delta^6}{K^2}\left((\lambda'\lambda)^2\Pi'\Pi+ \sum_{i,i',j,j'}P_{ij}^2|\Pi_i\lambda_i\Pi_j|P_{i'j'}^2| \Pi_{i'}\lambda_{i'}\Pi_{j'}|\sum_k|M_{jk}M_{j'k}|\right)\leq\\ \leq \frac{C\Delta^6}{K^2}\left((\lambda'\lambda)^2\Pi'\Pi+ (\Pi'\Pi)^2\lambda'\lambda\right)\leq \frac{C\Delta^6(\Pi'\Pi)^3}{K^3}\to 0.
\end{align*}
For the first inequality, we apply Assumption~\ref{ass: errors} and bound (\ref{eq: bound on Pij}).  Then we use Cauchy-Schwarz inequality for the first summand: $\left(\sum_iP_{ij}^2|\Pi_i||\lambda_i|\right)^2\leq\Pi'\Pi\lambda'\lambda$.  For the second summand, we apply
Lemma S1.1 (ii) and Lemma S1.3 (c). Finally, we apply Lemma S2.1 and S2.2 to get $\Delta^2A_2\to^p 0$ and $\Delta A_1\to^p0$. $\Box$

\paragraph{Proof of Theorem \ref{thm:AR power1}.}
The infeasible version of AR statistics under $\beta=\beta_0+\Delta$ is:
\begin{align}
 & \frac{1}{\sqrt{K}\sqrt{\Phi}}\sum_{i}\sum_{j\neq i}P_{ij}e_{i}(\beta_{0})e_{j}(\beta_{0})\nonumber \\
= & \frac{\Delta^{2}}{\sqrt{K}\sqrt{\Phi}}\sum_{i}\sum_{j\neq i}P_{ij}\Pi_i\Pi_j +\frac{2\Delta}{\sqrt{K}\sqrt\Phi}\sum_{i}\left(\sum_{j\neq i}P_{ij}\Pi_j\right)\eta_{i}+\frac{1}{\sqrt{K}\sqrt{\Phi}}\sum_{i}\sum_{j\neq i}P_{ij}\eta_{i}\eta_{j}. \label{eq: AR numerator}
\end{align}
The first term in (\ref{eq: AR numerator}) is deterministic and equals to $\Delta^2\frac{\mu^2}{\sqrt{K}\sqrt{\Phi}}$. The second term has mean zero and
 variance
$$
\frac{\Delta^{2}}{K\Phi}\sum_{i}\left(\sum_{j\neq i}P_{ij}\Pi_j\right)^2Var(\eta_i)\leq \frac{Cc^{2}}{K\Phi}\sum_{i}w_i^2\leq \frac{C\Pi'\Pi}{K}\to 0.
$$
Here we used that variance of $\eta_i$ is bounded by Assumption \ref{ass: errors}, $\sum_{j\neq i}P_{ij}\Pi_i=w_i$, and the final  bound is proven in  Lemma S1.4. Thus, the second term converges to zero in probability uniformly over $|\Delta|^2\leq c$. The third term in (\ref{eq: AR numerator}) is  asymptotically standard normal  due to Lemma \ref{lemma: CLT}.
Finally, we notice that
$$
AR(\beta_0)=\sqrt{\frac{\Phi}{\widehat\Phi}}\frac{1}{\sqrt{K}\sqrt{\Phi}}\sum_{i}\sum_{j\neq i}P_{ij}e_{i}(\beta_{0})e_{j}(\beta_{0}),
$$
and apply Theorem \ref{thm: consistency of variance alternative}. This finishes the proof of statement (\ref{eq: them about local power}).

Now consider the case when $\frac{\mu^2}{\sqrt{K}\sqrt{\Phi}}\to\infty$ and $\Delta\neq 0$ is fixed. Above we proved  that
$$
\frac{1}{\sqrt{K}\sqrt{\Phi}}\sum_{i}\sum_{j\neq i}P_{ij}e_{i}(\beta_{0})e_{j}(\beta_{0})=\frac{\mu^2}{\sqrt{K}\sqrt{\Phi}}\Delta^2+o_p(1)+O_p(1).
$$
Finally, Theorem \ref{thm: consistency of variance alternative} implies that  $\frac{\widehat\Phi}{\Phi}\to^p 1$.  As a result, we have $AR(\beta_0)\to^p\infty$ when $\frac{\mu^2}{\sqrt{K}\sqrt{\Phi}}\to\infty$ and $\Delta\neq 0$ is fixed. This lead to rejection probability converging to 1. $\Box$

\paragraph{Proof of Theorem \ref{thm: first stage F}.}
Denote
$$
Q=\left(
      Q_{ee},
    Q_{Xe} ,
    Q_{XX}
\right)^\prime=\frac{1}{\sqrt{K}}\sum_{i=1}^N\sum_{j\neq i}P_{ij}
\left(
    e_ie_j,
    X_ie_j ,
    X_iX_j
\right)^\prime.
$$
Lemma A2 in Chao et al. (2012) states  that for  any fixed $3\times 1$ vector $a$ we have
$
(a'\Sigma a)^{-1/2}\left(    Q_{ee},
    Q_{Xe} ,
    Q_{XX} -\frac{\mu^{2}}{\sqrt{K}}\right)a \Rightarrow N(0,1)$. According to Cram\'er-Wold theorem, this implies that $$
\Sigma^{-1/2}\left(    Q_{ee},
    Q_{Xe} ,
    Q_{XX} -\frac{\mu^{2}}{\sqrt{K}}\right) \Rightarrow N(0,I_3)$$  where
$\Sigma$
is the asymptotic covariance matrix  of $Q$, with some of its elements written below:
\begin{align}
\Psi & =\frac{1}{K} \sum_{i=1}^{N} \sum_{j\neq i} P_{ij}^{2} \gamma_i \gamma_j + \frac{1}{K} \sum_{i=1}^{N} \sum_{j\neq i} P_{ij}^{2} \sigma_i^2 \varsigma_j^2 +\frac{1}{K} \sum_{i=1}^{N}(\sum_{j\neq i}P_{ij}\Pi_j)^{2}\sigma_i^2=AVar(Q_{Xe}) , \notag \\
\Upsilon & =\frac{2}{K}\sum_{i=1}^{N}\sum_{j\neq i}P_{ij}^{2}\varsigma_i^2 \varsigma_j^2+\frac{4}{K}\sum_{i=1}^{N}\varsigma_i^2(\sum_{j\neq i}P_{ij}\Pi_j)^{2}=AVar(Q_{XX}),\label{eq: Upsilon}\\
\tau & =\frac{2}{K}\sum_{i=1}^{N}\sum_{j\neq i}P_{ij}^{2}\varsigma_i^2\gamma_j+\frac{2}{K}\sum_{i=1}^{N}\gamma_i(\sum_{j\neq i}P_{ij}\Pi_j)^{2} =ACov(Q_{Xe},Q_{XX}),~~~ \varrho =\frac{\tau}{\sqrt{\Psi}\sqrt{\Upsilon}}.\notag
\end{align}
where $\sigma_i^2=Var(e_i), \varsigma_i^2=Var(v_i)$, $\gamma_i=cov(e_i,v_i)$.
Note
that $\widehat{e}_{i}=Y_{i}-X_{i}\widehat{\beta}_{JIVE}=e_{i}-X_{i}(\widehat{\beta}_{JIVE}-\beta)$
and $(\widehat{\beta}_{JIVE}-\beta_{0})=Q_{Xe}/Q_{XX}$.
Thus,
\begin{align*}
Wald(\beta_0) & =\frac{Q_{Xe}^{2}}{\sum_{i=1}^{N}\left(\sum_{j\neq i}P_{ij}X_{j}\right)^2 \frac{\widehat{e}_{i}M_i\widehat{e}}{M_{ii}}+\sum_{i=1}^{N}\sum_{j\neq i}\widetilde{ P}_{ij}^{2}M_iX\widehat{e}_{i}M_jX\widehat{e}_{j}},
\end{align*} where the denominator expands to
\begin{align*}
&\sum_{i=1}^{N}\left(\sum_{j\neq i}P_{ij}X_{j}\right)^2\frac{\widehat{e}_{i}M_i\widehat{e}}{M_{ii}}+ \sum_{i=1}^{N}\sum_{j\neq i}\widetilde{ P}_{ij}^{2}M_iX\widehat{e}_{i}M_jX\widehat{e}_{j}=\\
=&\left\{\sum_{i=1}^{N}\left(\sum_{j\neq i}P_{ij}X_{j}\right)^2\frac{e_{i}M_ie}{M_{ii}}+ \sum_{i=1}^{N}\sum_{j\neq i}\widetilde{ P}_{ij}^{2}M_iXe_{i}M_jXe_{j}\right\} - \\
-&  \frac{Q_{Xe}}{Q_{XX}}\left\{\sum_{i=1}^{N}\left(\sum_{j\neq i}P_{ij}X_{j}\right)^2\left(\frac{e_{i}M_iX}{M_{ii}}+ \frac{X_{i}M_ie}{M_{ii}}\right)+2\sum_{i=1}^{N}\sum_{j\neq i}\widetilde{ P}_{ij}^{2}M_iXe_{i}M_jXX_{j}\right\}+\\
+& \frac{Q_{Xe}^2}{Q_{XX}^2}\left\{\sum_{i=1}^{N}\left(\sum_{j\neq i}P_{ij}X_{j}\right)^2\frac{X_{i}M_iX}{M_{ii}}+\sum_{i=1}^{N}\sum_{j\neq i}\widetilde{ P}_{ij}^{2}M_iXX_{i}M_jXX_{j}\right\}.
\end{align*}
Applying Lemma S3.1 from the Supplementary Appendix to the expanded expression of the denominator, we show the terms appearing in the braces converge to $\Psi, 2\tau$ and $\Upsilon$ respectively.
Then
\begin{align*}
Wald(\beta_0)  =\frac{Q_{Xe}^{2}}{\Psi-2\frac{Q_{Xe}}{Q_{XX}}\tau+\frac{Q_{Xe}^2}{Q_{XX}^2}\Upsilon}(1+o_p(1))= \frac{Q_{Xe}^{2}/\Psi}{1-2\frac{Q_{Xe}/\sqrt{\Psi}}{Q_{XX}/\sqrt{\Upsilon}}\varrho+\frac{Q_{Xe}^2}{Q_{XX}^2}\frac{\Upsilon}{\Psi}}(1+o_p(1)).
\end{align*}
Lemmas \ref{lem: consistency of variance null} and \ref{lem: on estimation under alternative} applied to $\widehat{\Upsilon}$ with $\xi_i=(v_i,v_i,v_i,v_i)'$ and $\Delta=1$ give $
\widetilde{F}  =\frac{Q_{XX} }{\sqrt{\Upsilon}}(1+o_p(1)).
$
Thus, the statement of Theorem \ref{thm: first stage F} holds where we denote  $\left(
\xi,\nu-\frac{\mu^{2}}{\sqrt{K}\sqrt{\Upsilon}}\right)$ to be the Gaussian limit of $(\frac{Q_{Xe}}{\sqrt{\Psi}},\frac{Q_{XX}}{\sqrt{\Upsilon}}-\frac{\mu^{2}}{\sqrt{K}\sqrt{\Upsilon}})$. $\Box$

\paragraph{Proof of Corollary \ref{cor: only one}.} Denote $S=\frac{\mu^{2}}{\sqrt{K}\sqrt{\Upsilon}}$. If $S>2.5$ then due to Theorem \ref{thm: first stage F}:
$$
\mathbb{P}_S\{\widetilde{F}>4.14\text{ and }Wald(\beta_{0})\geq\chi_{1,0.95}^{2}\}\leq \mathbb{P}_S\{Wald(\beta_{0})\geq\chi_{1,0.95}^{2}\}\leq 0.10.
$$
If $S\leq 2.5$ then due to the asymptotic gaussianity of $\widetilde{F}$:
\begin{align*}
\mathbb{P}_S\{\widetilde{F}>4.14\text{ and }Wald(\beta_{0})\geq\chi_{1,0.95}^{2}\}\leq \mathbb{P}_S\{\widetilde{F}>4.14\}\leq 0.05.
\end{align*}
Finally, for any $S>0$:
\begin{align*}
 \mathbb{P}_S\left\{ H_{0}\text{ is rejected }\right\} = \mathbb{P}\{\widetilde{F}>4.14\text{ and }Wald(\beta_{0})\geq\chi_{1,0.95}^{2}\}+\\
 +\mathbb{P}\{\widetilde{F}\leq 4.14\text{ and }AR(\beta_{0})\geq z_{0.95} \}\leq 0.10+ \mathbb{P}\{AR(\beta_{0})\geq z_{0.95} \}\leq 0.15.
\end{align*}
\subsection{Simulation Details}\label{appendix - simulation details}
\paragraph{For results reported in Section \ref{section - simulation}.} To create many instruments, we interact QOB dummies with dummies for
year of birth (YOB) and place (state) of birth (POB).  Interacting three QOB dummies with nine YOB and 50 POB dummies
generates 180 excluded instruments. The excluded instruments
are
\begin{align*}
Z_{i}=((\mathbf{1}\{Q_{i}=q,C_{i}=c\})_{q\in\{2,3,4\},c\in\{31,\dots,39\}}',\mathbf{1}\{Q_{i}=q,P_{i}=p\})_{q\in\{2,3,4\},p\in\{50\ \text{states}\}}')',
\end{align*}
where $Q_{i},C_{i},P_{i}$ are $i$'s QOB, YOB and POB respectively.
Note,  that $Z_{i}$ are not group instruments in the strict sense as they are
not mutually exclusive. We exclude instruments with $\sum_{i=1}^{N}Z_{ij}<5$
to satisfy the balanced instruments assumption (Assumption \ref{ass: projection matrix}).

To increase the amount of omitted variable bias, we follow Angrist
and Frandsen (2019) by taking the LIML model as the ground truth,
where the outcome variable is $Y_{i}$ (income), the endogenous variable
$X_{i}$ (highest grade completed) is instrumented by $Z_{i}$ and
the control variables are a full set of POB-by-YOB interactions. Specifically,
starting with the full 1980 census sample, we compute the average
$X_{i}$ in each QOB-YOB-POB cell $\bar{s}(q,c,p)$ . We then estimate
LIML and retain $\hat{y}(c,p)$, the second-stage fitted value after
subtracting $\hat{\beta}_{LIML}X_{i}$ where $\hat{\beta}_{LIML}$
is the LIML estimate of the returns to schooling. We also retain the
variance of LIML residuals $\omega(Q_{i},C_{i},P_{i})$ to mimic the
heteroskedasticity.

The simulation model we consider is then
\begin{align*}
\tilde{y}_{i} & =\bar{y}+0.1\tilde{s}_{i}+\omega(Q_{i},C_{i},P_{i})(\nu_{i}+\kappa_{2}\epsilon_{i})\\
\tilde{s}_{i} & \sim Poisson(\mu_{i}),
\end{align*}
for independent standard normal $\nu_{i}$ and $\epsilon_{i}$. Here
$\bar{y}=\frac{1}{N}\sum_{i}\hat{y}(C_{i},P_{i})$ and $\mu_{i}=\max\{1,\gamma_{0}+\gamma_{Z}'Z_{i}+\kappa_{1}\nu_{i}\}$
where $\gamma_{0}+\gamma_{Z}'Z_{i}$ is the projection of $\bar{s}(Q_{i},C_{i},P_{i})$
onto a constant and $Z_{i}$.  We set $\kappa_{1}=1.7$ and $\kappa_2 = 0.1$ following Angrist and Frandsen (2019).  The first stage is therefore nonlinear non-linear
in $Z_{i}$ as $\mu_{i}$ is a censored normal random variable.  The first stage error is heteroskedastic and the theoretical variance can be derived analytically.

\paragraph{For results reported in Section~\ref{subsection - power}.}
The DGP is given by a homoscedastic linear IV model (\ref{eq: iv model}) with a linear first stage $\Pi_i=\Pi'Z_i$. The instruments are $K=40$ group indicators, where the sample is divided into equal groups.  The sample size is $N=200$. The error terms are generated i.i.d. as $
\left(\begin{array}{c}
e_{i}\\
v_{i}
\end{array}\right)\sim \mathcal{N}\left(\left(\begin{array}{c}
0\\
0
\end{array}\right),\left(\begin{array}{cc}
1 & \rho\\
\rho & 1
\end{array}\right)\right)
$
with $\rho=0.2$. We simulate a sparse first stage by setting one large coefficient $\pi_K=2$ and   $\pi_{k}=0.001$ for all $k<K$. The dense first stage has homogeneous first stage coefficients  $\pi_{k}=0.316$ for all $k=1,\dots,K$. Identification strength is held the same at $\frac{\mu^2}{\sqrt{K}}=2.5$ for both settings. The results are reported in Figure~\ref{fig:power dense}.

\end{document}